\documentclass{pasa}%

\title[Formation of prestellar cores via non-isothermal gas fragmentation]{Formation of prestellar cores via non-isothermal gas fragmentation}
\author[Anathpindika, S]{Anathpindika, S$^1$ \thanks{sumed$\_$k@yahoo.co.in; sumedh@physics.iisc.ernet.in}\\
\affil{$^1$Indian Institute of Science}}%
\jid{PASA}
\doi{10.1017/pas.\the\year.xxx}
\jyear{\the\year}

\begin{document}%
\begin{abstract}
 Sheet-like clouds are common in turbulent gas and perhaps form via collisions between turbulent gas flows. Having examined the evolution of an isothermal shocked slab in an earlier contribution, in this work we follow the evolution of a sheet-like cloud confined by (thermal)pressure and gas in it is allowed to cool. The extant purpose of this endeavour is to study the  early phases of core-formation.  The observed evolution of this cloud supports the conjecture that molecular clouds themselves are three-phase media (comprising viz. a stable cold and warm medium, and a third thermally unstable medium), though it appears, clouds may evolve in this manner irrespective of whether they are gravitationally bound. We report, this sheet fragments initially due to the growth of the thermal instability and some fragments are elongated, filament-like. Subsequently, relatively large fragments become gravitationally unstable and sub-fragment into smaller cores. The formation of cores appears to be a three stage process : first, growth of the thermal instability leads to rapid fragmentation of the slab; second, relatively small fragments acquire mass via gas-accretion and/or merger and third, sufficiently massive fragments become susceptible to the gravitational instability and sub-fragment to form smaller cores. We investigate typical properties of clumps (and smaller cores) resulting from this fragmentation process. Findings of this work support the suggestion that the weak velocity field usually observed in dense clumps and smaller cores is likely seeded by the growth of dynamic instabilities. Simulations were performed using the smooth particle hydrodynamics algorithm.
\end{abstract}
\maketitle%
\section{INTRODUCTION }
\label{sec:intro}
Over the last two decades star-forming clouds({\small SFC}s) have been studied in ever increasing detail with telescopes on the ground like the {\small JCMT} (e.g. Ward-Thompson \emph{et al.} 2007), and those in space like Spitzer (Kirk \emph{et al.} 2007), and Herschel (e.g. Andr{\' e} \emph{et al.} 2010). While  improving our understanding about the initial conditions for the formation of stars on the one hand, these studies, on the other, have also posed a few challenging questions. Consequently, recent years have seen considerable investment of observational and computational efforts in this subject. Modulo the effect of magnetic field, which is also believed to play a significant role in the formation of stars, purely hydrodynamic models  have, to a large extent, successfully reconciled the physical conditions found in typical star-forming clouds (for e.g., see review by Hennebelle \& Falgarone 2012). 

Presently we face  at least two major questions that demand investigation : (i) the formation and evolution of prestellar cores in putative star-forming clouds, and (ii) the relationship, if any, between the distribution of stellar masses, as described by the stellar initial mass function, and that for prestellar cores\footnote{In the rest of this contribution \emph{prestellar cores} will be simply referred to as cores.} (e.g. Alves \emph{et al.} 2007; Nutter \& Ward-Thompson 2007; Enoch \emph{et al.} 2008; Anathpindika 2011). At the moment though, our interest lies in studying the formation of cores and the development of their physical properties. The question has evoked considerable interest among a number of authors and there has been a growing consensus on the formation of cores via fragmentation of dense clumps within turbulent {\small MC}s. This model envisages a competition between self-gravity and turbulence (see e.g., Hartmann 2002; Klessen \emph{et al.} 2005; Bonnell \emph{et al.} 2006; Heitsch \emph{et al.} 2008b; Schmidt \emph{et al.} 2010; Price \emph{et al.} 2011; Federrath \& Klessen 2012 and Anathpindika 2013). The cascading nature of turbulence makes it crucial in the evolutionary cycle of {\small MC}s. On the sub-parsec scales it induces the formation of local compression that may self-gravitate if the Jeans condition is satisfied (see e.g. Mac Low \& Klessen 2004; Audit \& Hennebelle 2010; Bonnell \emph{et al.} 2013).

Turbulence and the fragmentation of turbulent gas is scale-free which makes it plausible that clouds and cores form in identical processes. Indeed, sub-fragmentation within larger clouds has been reported and a more recent example being the starless cloud, {\small IRDC}18310-4. The length-scale of internal fragmentation 
within this cloud was found to be consistent with the local thermal Jeans length (Beuther \emph{et al.} 2013). A number of similar examples have been reported and now form part of the voluminous literature on this subject. Clouds and smaller clumps that form in this fashion could be bound, although  evidence from numerical simulations suggests otherwise. It has been suggested, fragmentation of shocked bodies though rapid, is more likely to produce unbound fragments, confined by external pressure and may merge to form larger clumps. These fragments are unlike those that form under the assumption of isothermality in which case they are likely to be relatively short-lived on account of their propensity towards re-expansion due to insufficient confining pressure (see e.g. V{\' a}zquez-Semadeni \emph{et al.} 2007; Heitsch \emph{et al.} 2008 b; Anathpindika 2009).

Surveys of {\small MC}s have revealed populations of both bound, as well as unbound cores, either of which could remain starless. Despite their relatively high levels of turbulence, some  do indeed have a population of cores, most of which are unbound and therefore starless (see e.g., Hatchell \emph{et al.} 2005 for the Perseus {\small MC}; Hily-Blant \& Falgarone 2009 for Polaris Flare; Roman-Duval \emph{et al.} 2010 for {\small MC}s in the Galactic ring survey; Ragan \emph{et al.} 2012 for a number of {\small IRDCs}\footnote{The infrared dark clouds(IRDCs) are among the densest clouds and appear dark against a bright IR background.} and Ripple \emph{et al.} 2013 for the Orion MC). Starless cores are believed to be supported by turbulence, and perhaps magnetic field. The occurrence of a weak velocity field within these cores is attributed to the cascading nature of turbulence as it trickles down from large spatial scales to smaller ones (see e.g. Goodman \emph{et al.} 1998; Vazquez-Semadeni \emph{et al.} 2003; Federrath \emph{et al.} 2010; Federrath 2013). Broadening of optically thick emission lines  from  molecular tracers such as {\small CO}, {\small NH$_{3}$}, {\small H$_{2}$O} and {\small SiO} is a good tracer of the turbulent nature of clouds on the sub-parsec scale, i.e. on the scale of cores.

More stable fragments forming out of weakly turbulent gas successfully reproduce crucial properties such as the mass-spectrum and further fragmentation produces smaller prestellar cores, as argued analytically by Hennebelle \& Chabrier (2008), and Anathpindika (2013). Numerical simulations on the subject also demonstrate the formation of an extensive network of dense filaments within turbulent gas (e.g. Price \& Bate 2008, 2009; Bate 2009; Federrath \& Klessen 2012). Thus we would like to see first, how sheet-like clouds are converted into filaments, second, how cores form in these filaments, and third, the occurrence of a relatively weak velocity-field on the scale of prestellar cores. The problem has been attempted both, analytically as well as numerically. For instance, stability analysis of pressure-confined and/or unconfined isothermal sheet-like clouds has shown that the fastest growing unstable mode is on the order of a fraction of the slab-thickness (Elmegreen \& Elmegreen 1978; Lubow \& Pringle 1993).

Pressure-confined slabs in the limiting case of confinement due to thermal-pressure or those due to shocks, show mutually different behaviour. For example, numerical work by  Inutsuka \& Miyama (1997) and Anathpindika (2009) demonstrated that a shocked slab, after its formation, soon lost thermal support due to internal dissipation as fluid layers excited by hydrodynamic instabilities sheared against each other. Unsupported clouds then lost thermal support  and collapsed to the central plane before expanding out laterally to form a filament. Slabs confined by thermal-pressure
of course, behave differently as they break-up into smaller fragments. The susceptibility of these sheets to changes in external pressure was first discussed by Hunter (1979), Whitworth (1981) and Hunter \& Fleck (1982), and more recently, by Heyer \emph{et al.} (2009). Similarly, the importance of cooling in shocked gas-bodies was demonstrated by Tohline \emph{et al.} (1987). According to the hypothesis, efficient cooling of a pressure-compressed gas body raised its propensity towards fragmentation so that one/many of these fragments could eventually spawn stars.  Pressure-confined isothermal gas slabs were also examined by Whitworth \emph{et al.} (1994), Yamada \& Nishi (1998) and Clarke (1999). 

It is evident that strength of the external pressure could possibly control the evolutionary cycle of a confined cloud. In the present work we propose to examine the behaviour of an initially isothermal sheet-like cloud subject to strong pressure confinement, but not a shock, and is allowed to cool over the course of its evolution. In particular, we are interested in - \textbf{(a)} examining the interplay between the gravitational and the thermal instability during the process of gas fragmentation, \textbf{(b)} examine the behaviour of gas in the warm and cold phase and finally, \textbf{(c)} test the consistency between the properties of cores derived here and those reported for field cores. This contribution is divided as follows : the initial conditions for the simulations and the numerical algorithm used are discussed in \S 2. Then in \S 3, the evolution of the pressure-confined slab is discussed and useful diagnostics are presented in \S 4 and \S 5. We conclude in \S 6. 

\begin{figure}
\centering
  \includegraphics[angle=0,width=0.5\textwidth]{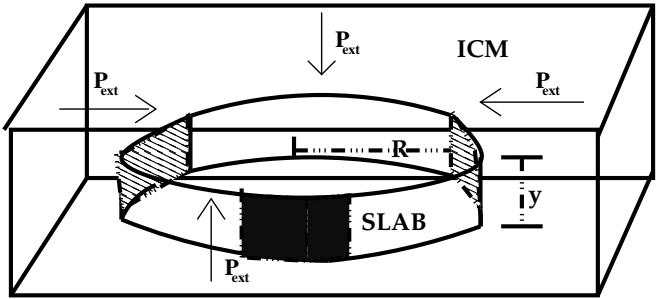}
  \caption{A sketch of the gas-slab of radius, $R$, and height, $y$,  confined by the ICM has been shown in this cartoon. The small ICM exerts pressure, $p_{ext}$, externally on all faces of the slab as demonstrated by the direction of arrows. }
\end{figure}

\begin{table*}
\begin{minipage}[t]{\textwidth}
 \caption{Physical parameters for the two cases.}
\label{table:1}      
\centering          
  \begin{tabular}{l l l l l l l l l l l l}
  \hline
    Serial     & $\bar{n}_{gas}$\footnote{Number density for gas and the {\small ICM} listed in columns 2 \& 3 in units of [cm$^{-3}$].} & $\bar{n}_{icm}$ & $\frac{T_{gas}}{[K]}$ & $\frac{T_{icm}}{[K]}$  & $\Big(\frac{P_{therm}\footnote{Components of pressure listed in columns 6, 7 \& 8 in units of $10^{4}\times$[$\mathrm{K\ cm}^{-3}$].}}{k_{B}}\Big)$ & $\Big(\frac{P_{grav}}{k_{B}}\Big)$ & $\Big(\frac{P_{ext}}{k_{B}}\Big)$ & $\frac{h_{0}}{[pc]}$\footnote{Scale-height for the slab} & $\frac{N_{tot}\footnote{Total number of particles including those representing the {\small ICM}. }}{\times 10^{5}}$ & $\frac{h_{avg}}{[\mathrm{pc}]}$\footnote{Average smoothing length for gas particles.} & $\frac{M_{gas}}{M_{Jeans}}$ \footnote{$M_{Jeans}$ is the thermal Jeans mass calculated for the initial gas temperature, $T_{gas}$.}\\
 \hline
1 & 2.5$\times 10^{2}$ & 125 & 40 & 131 & 1.0 & 0.63662 & 1.63662 & 0.572  & 6.4 & 0.044 & 1.4 \\
2 & 2.5$\times 10^{2}$ & 125 & 100 & 328 & 2.5 & 1.592 & 4.092 & 0.905  & 7.21 & 0.052 & 0.6 \\
3 & 2.5$\times 10^{2}$ & 125 & 150 & 491 & 3.75 & 2.388 & 6.138 & 1.108 & 7.61 & 0.055 & 0.36\\
\hline
\end{tabular}
\end{minipage}
\end{table*}

\section{Initial conditions}
The set-up for our simulations is simple and consists of a uniform-density slab that is marginally gravitationally super-critical in case 1 (slab mass approximately 1.4 times the initial thermal Jeans mass), and gravitationally sub-critical (slab mass only a fraction of the initial thermal Jeans mass) in the remaining two cases (see Table 1). The slab is circular with radius, $R$, and is in pressure-equilibrium with the inter-cloud medium ({\small ICM}), that confines it. Note that particles in the slab and the {\small ICM} were so distributed that the number-density of particles across the slab-{\small ICM} interface is constant.  The entire assembly was placed in a box, but without periodic-boundary conditions, meant only to prevent the {\small ICM} from diffusing away so that particles escaping from one face of the box re-enter from the opposite face.  That {\small SPH} particles representing the {\small ICM} are mobile and exert only hydrodynamic force on normal gas particles would serve as a useful reminder to our readers at this point. The schematic sketch in Fig. 1 illustrates this initial set-up for the simulations. Gas within the slab is composed of the usual cosmic mixture maintained initially at temperature, $T_{gas}$. We set $\mu$ = $4\times 10^{-24}$gms, the mean molecular mass of the gas particles.  

The slab was characterised by radius, $R$ = 2.5 pc, uniform initial number density, $\bar{n}$ = 250 cm$^{-3}$, gas temperature, $T_{gas}$, and thickness, $y\ =\ 2h_{0}$, where the scale-height,  $h_{0}$, for the slab has been calculated below. Values of the relevant physical parameters have been listed in Table 1. For a slab in pressure equilibrium, the pressure exerted by the {\small ICM}, $P_{ext}$, balances the thermal pressure, $P_{therm}$, and the gravitational pressure, $P_{grav}$, within the slab so that
\begin{equation}
P_{ext} = P_{int} + P_{grav} = P_{int} + G\Sigma_{g}^{2}
\end{equation}
(e.g. Whitworth et al. 1994), where the gas surface-density, 
$\Sigma_{g} = \bar{n}\mu h_{0}$. The gas temperature, $T_{gas}$, corresponds to sound-speed $a$, so that the scale height for the slab, $h_{0}\ =\ \frac{a}{\sqrt{2\pi G\bar{n}\mu}}$ (Ledoux 1951; Lubow \& Pringle 2001). The length of the fastest growing mode in a cooling medium is,
\begin{equation}
\lambda_{fast} = \Big(\frac{\gamma_{e}\pi a^{2}}{\gamma G\mu\bar{n}}\Big)^{1/2}
\end{equation}
(V{\' a}zquez-Semadeni \emph{et al.} 1996); $\gamma_{e}$, is the effective polytropic exponent, i.e. slope  of the cooling curve shown in Fig. 2, which for our choice of the initial gas density, $\bar{n}$, turns out to be $\sim$ 0.9 where as, $\gamma$ = 5/3, is the adiabatic gas constant. The equilibrium pressure at density, $\bar{n}$, according to the cooling curve is $~7500$ K cm$^{-3}$. We perform three realisations with different choice of the initial gas temperature, and therefore, thermal pressure, $P_{therm}$, away from its equilibrium magnitude as can be seen from Table 1.

\begin{figure}
  \vspace*{2pt}
  \includegraphics[angle=270,width=0.5\textwidth]{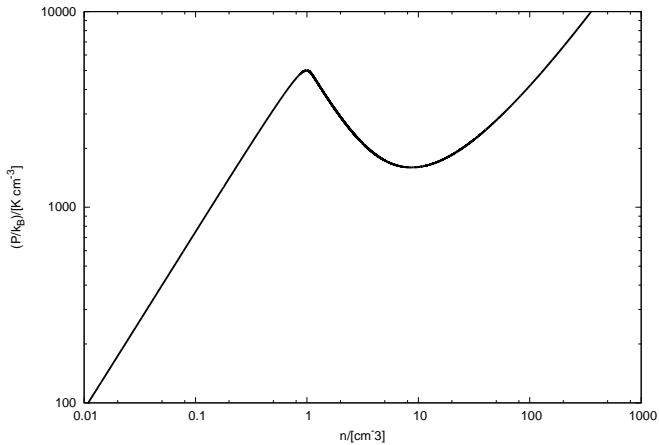}
  \caption{The thermal equilibrium curve showing a plot of gas pressure against its density. }
\end{figure}

\begin{figure*}
  \vspace*{2pt}
  \includegraphics[angle=270,width=\textwidth]{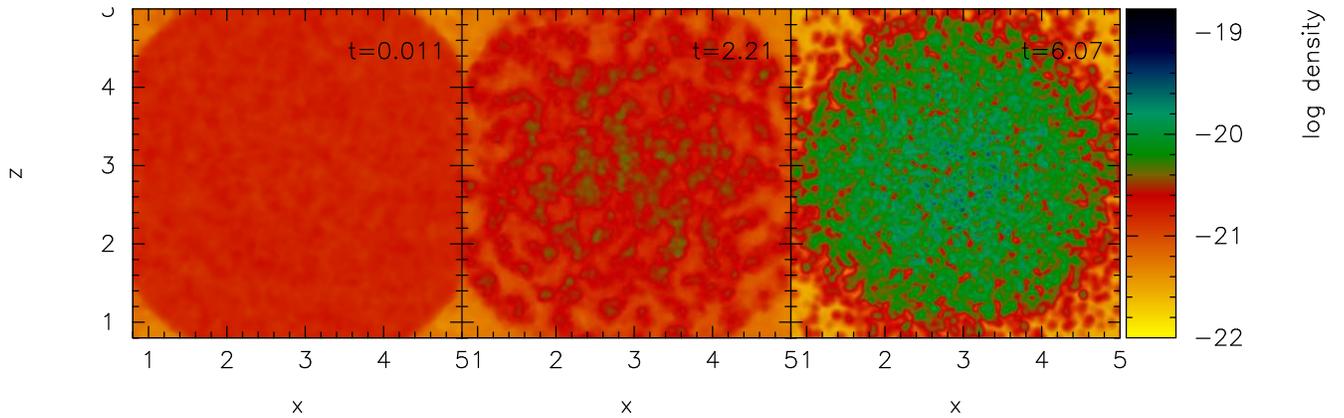}
  \caption{Plot showing rendered density images of a projection of the mid-plane of the slab in case 2. Time measured in Myrs has been marked on the top right-hand corner and spatial coordinates in units of parsecs.}
\end{figure*}

\subsection{Numerical algorithm}
 Numerical simulations were performed using the well-tested Smoothed particle hydrodynamics ({\small SPH}) code {\small SEREN} in its energy and momentum conserving formalism with an adaptive gravitational softening length (Hubber \emph{et al.} 2011). Simulations were developed with the standard cubic-spline kernel with each {\small SPH} particle having approximately 50 neighbours. We note, some concerns have been reported in the literature (e.g. Hobbs \emph{et al.} 2013), about the suitability of the classical {\small SPH} algorithm to investigate the fragmentation of a cooling gas on account of its inability to handle the mixing between hot and cold gas particles. It would be appropriate to point out that the said problem reported by these authors is in the extreme environment of Galactic accretion where gas temperature is at least five orders of magnitude higher than that in the present work so that the supposed difficulty with the mixing between two species of {\small SPH} particles could present a problem. This is unlikely in the present work where we are working with gas at a temperature not greater than a few hundred Kelvin. On the contrary, {\small SPH} has been routinely used to study problems such as the formation of Giant molecular clouds (GMCs) in the galactic disk via fragmentation of gas that is allowed to cool (e.g. Dobbs et al. 2012; Bonnell \emph{et al.} 2013), a problem that is of a similar nature to the one at hands in this work, but of course, on a much larger spatial scale. Nevertheless, in order to mitigate any such eventuality we use the prescription of artificial conductivity suggested by Price (2008). The Riemann artificial viscosity prescribed by Monaghan (1997) is used in these calculations with the corresponding viscosity parameter, $\alpha$ = 0.5. 

The average initial smoothing length of a particle, $h_{avg}$, is defined as 
\begin{displaymath}
h_{avg}^{3}\sim \frac{3V_{gas}}{32\pi}\Big(\frac{N_{neibs}=50}{N_{gas}}\Big);
\end{displaymath}
 $V_{gas}$ is the volume of the slab. Each simulation in this work was developed using 3.83$\times$10$^{5}$ gas particles.
For the magnitude of the average smoothing length, $h_{avg}$, listed in column 11 of Table 1, the resolution criterion defined by Hubber \emph{et al.} (2006), the {\small SPH} equivalent of the well-known Truelove condition (Truelove et al. 1997), is readily satisfied down to the typical size of a prestellar core ($\sim$ 0.2 pc), at $10^{4}$ cm$^{-3}$ and 10 K. This is good enough to study the formation of filaments and clumps in the post-fragmentation gas-slab. Also, note that the average smoothing length, $h_{avg}$, is at least an order of magnitude smaller than the slab-height so that there is sufficient spatial resolution along this direction as well.
However, although this choice of resolution is sufficient to prevent artificial fragmentation on the scale of a typical prestellar core in this work (at $l\sim$0.2 pc), it is inadequate to achieve convergence in the calculation of physical quantities such as energy and momentum of the gas on this scale (see Federrath \emph{et al.} 2014 \& 2011). To this end these authors have suggested a much more stringent criterion, $\frac{l}{h_{avg}}\gtrsim 30$, to be able to satisfactorily follow the dynamics on this scale. This adventure would of course, demand at least an order of magnitude increase in the number of gas particles. But we are not particularly concerned about this matter as it is not our intention to follow here the evolution of individual cores. Our immediate interest lies in demonstrating the initial stage of prestellar core-formation, i.e., the formation of dense clumps via fragmentation of gas that is allowed to cool. The average smoothing length, $h_{avg}$, in each of the three realisations is sufficiently small to satisfy the above mentioned resolution criterion at the length-scale of fragmentation, $\lambda_{fast}$, defined by Eqn. (2).

 In {\small SPH} the {\small ICM} is modelled with special type of particles that interact with normal gas particles only via the hydrodynamic force and are maintained at a fixed temperature, $T_{icm}$, throughout the entire length of each simulation. The temperature of ordinary gas particles, on the other hand, was determined by solving the equation for internal energy, modified
 suitably using the function below to account for cooling of the warm gas. 
Gas particles were allowed to cool by employing a parametric cooling function($\Lambda$), defined by Eqn. (3) below and plotted in Fig. 2. The curve can be readily identified as the thermal equilibrium curve for gas in the interstellar medium (e.g. Wolfire et al. 1995). We also assume a constant background source of heating defined by the heating function($\Gamma$), 
\begin{displaymath}
\Gamma = 2.0\times 10^{-26} \mathrm{erg}\ \mathrm{s}^{-1},  
\end{displaymath}
\begin{equation}
\frac{\Lambda(T)}{\Gamma} = 10^{7}\mathrm{exp}\Big(\frac{-1.184\times 10^{5}}{T + 1000}\Big) + \newline
1.4\times 10^{-2}\surd{T}\ \mathrm{exp}\Big(\frac{-92}{T}\Big)\ \mathrm{cm}^{3},
\end{equation}
(Koyama \& Inutsuka 2002; V{\' a}zquez-Semadeni \emph{et al.} 2007).


\begin{figure*}
  \vspace*{1pt}
  \mbox[\includegraphics[angle=270,width=\textwidth]{RUN2clumpmergedens.eps}
\includegraphics[angle=270,width=0.8\textwidth]{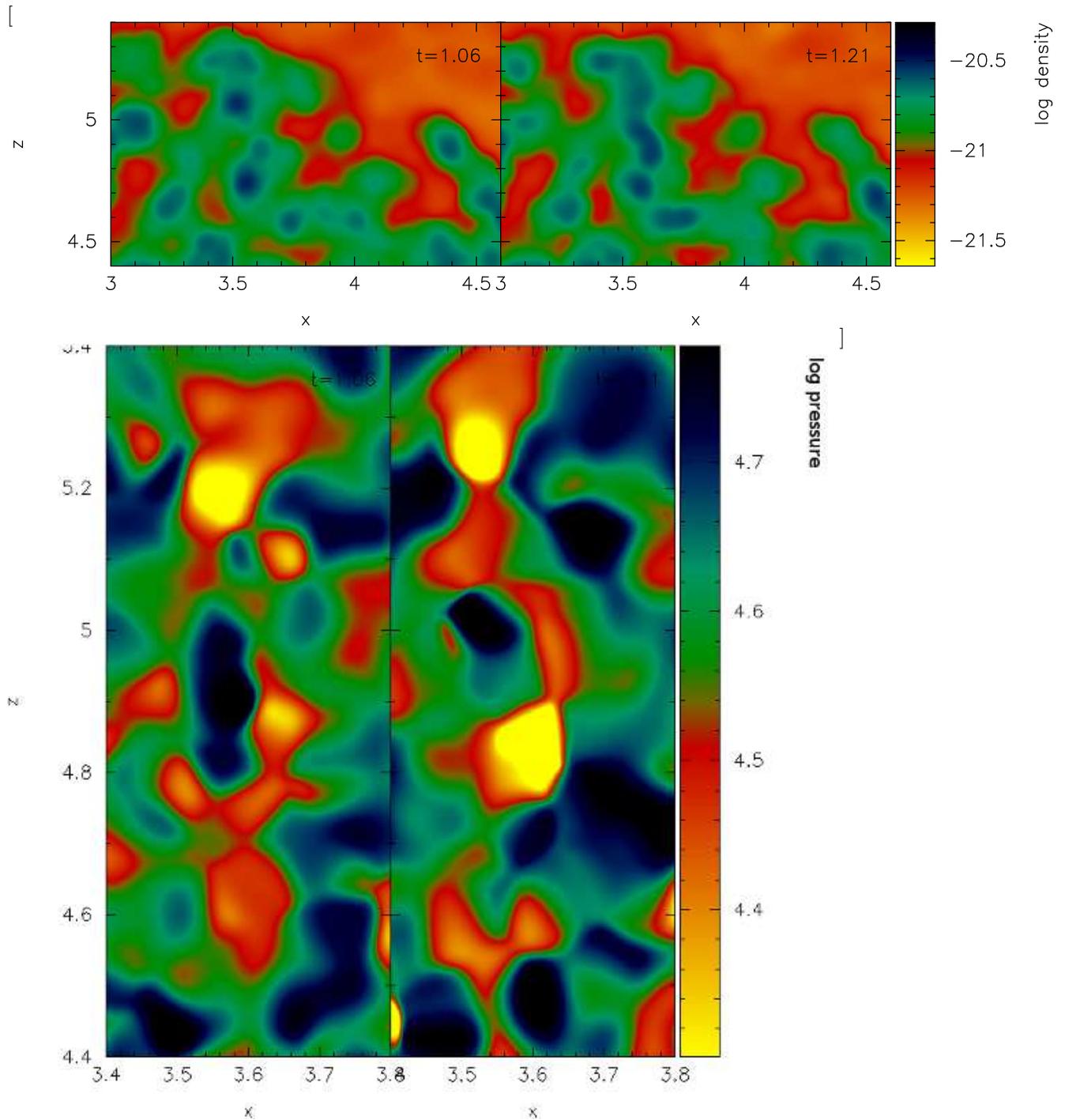}]
  \caption{Shown on the upper panel, in a set of images, is a small section of the fragmenting slab, fairly early in its evolution, for case 2. Of particular interest is the region between $x\in(3.5,3.7)$
 and $z\in(4.6,5.2)$, on these plots which shows the merger of smaller clumps leading to the formation of a single large clump that incidentally appears filament-like. Shown on the lower panel are rendered images of net pressure (in units of $\log$(K cm$^{-3}$)) over this region at the same epoch as that for pictures on the upper-panel. The merger of clumps to form a single contiguous object is reflected by contiguous nature of the gas-pressure over this region. As in Fig. 3, unit of time is Myrs and has been marked at the top right-hand corner of the upper panel. }
\end{figure*}

\textbf{\emph{Identifying gas clumps}} \\
We employed the robust {\small HOP} algorithm suggested by Eisenstein \& Hut (1998)  to identify clumps within the density field. The algorithm proceeds by first identifying the local density peaks in a given density field, followed by a neighbour search for each of the identified density peaks. For the purpose, we need three density thresholds : $\rho_{peak}$, $\rho_{saddle}$ and $\rho_{outer}$. An initial scan of the density field was carried out and {\small SPH} particles with density higher than, $\rho_{peak}\sim 10^{4} cm^{-3}$, were identified as density peaks, or in other words, the seed-particles for clumps. This choice of the density threshold is sufficiently large for the extant purpose of studying the onset of core-formation. The density at the edge of a fragment is denoted as, $\rho_{outer}$, and $\rho_{outer}\sim\ 0.2\rho_{peak}$. The density field was then re-scanned with this density threshold to identify immediate neighbours of the previously identified density peaks. Finally, two clumps were merged together to form a contiguous object if a particle and its nearest neighbour appeared in two different clumps and if the average density of these two particles turned out to be greater than $\rho_{saddle}\sim\ 0.1\rho_{peak}$. The radius of a clump was calculated as, $r_{clump} = \mathrm{max}(<\textbf{r}_{c} - \textbf{r}_{i}>)$, $i\in(1,N_{clump})$ and $c$, the identifier of the seed-particle for a clump. Finally, the mass of a clump was calculated as, $M_{clump}=\Sigma_{clump}r_{clump}^{2}$; $\Sigma_{clump}$, being the average surface density of a detected clump, and
\begin{displaymath}
\Sigma_{clump} \sim \frac{1}{N_{clump}}\sum_{p=1}^{N_{clump}}\rho_{p} h_{p}
\end{displaymath}
where, $h_{p}$ and, $\rho_{p}$, are respectively the smoothing length and density, of a particle belonging to a clump. The summation in the expression above is over the number of particles, $N_{clump}$, that constituted a clump. 

\section{Results} 
\subsection{Evolution of the pressure-confined slab}
Under confinement from external pressure, $p_{ext}$, the gas-slab fragments to produce  clumps, some of which are isolated, many other appear contiguous \footnote{Clumps that appear contiguous will hereafter be described as filamentary.}. In the absence of any external perturbations structure in the density field is seeded purely by numerical noise. While the slab in each of the three cases evolves in a mutually similar fashion, in case 1, where the slab was initially gravitationally super-critical, it also exhibits a tendency to contract in the radial direction. In the remaining two realisations the slab does not contract in the radial direction, but fragments, though the fragmentation-timescale successively increases in these  two realisations, viz. 2 and 3. Shown in Fig. 3 are rendered images of the mid-plane of the slab in case 2.  That the slab soon becomes flocculated can be seen from the picture on the central-panel of this figure. On the right-hand panel in this figure is the picture showing the fragmented slab at the time of terminating calculations in this realisation. At this epoch the slab is replete with filamentary structure and smaller cores appear to have begun forming along the length of some of these filaments. 

Rendered images in Fig. 3 also suggest that the fragmentation has probably been hierarchical, in the sense that we observe formation of larger, contiguous clumps that appear filament-like, which then sub-fragment to form smaller cores. However, we could not track the process further due to lack of adequate computational resources. 
Indeed, during the early stages of fragmentation some clumps do merge to form larger clumps. Shown on the upper panel of Fig. 4 for instance, is the rendered density plot of a small section of the fragmented slab. These pictures demonstrate the merger of clumps to form a single contiguous object that incidentally looks filament-like. But this is not to suggest that filamentary objects form exclusively via merger of smaller fragments. Observational evidence for such occurrence in molecular clouds has been reported;  star-formation in the Serpens molecular cloud, for instance, appears to have been triggered due to collision between two smaller clumps of molecular gas (e.g. Duarte-Cabral \emph{et. al.} 2010; Graves \emph{et al.} 2010). Furthermore, Andr{\' e} \emph{et al.}(2007) have estimated the collision time on the order of $\sim$1 Myr for cores in the Ophiuchus molecular cloud. More recently,  Tsitali \emph{et al.}(2014), have deduced an even shorter collision time on the  order of $\sim$0.7 Myr for cores in the Chamaeleon I North region.

Merger of fragments  has also been reported previously in numerical simulations (e.g. Banerjee \emph{et al.} 2009), and probably, is the result of clumps having lower thermal pressure relative to that in the medium between clumps, the inter-clump medium. Clumps floating in the tenuous inter-clump medium and confined by the pressure exerted by this medium may be thought of as analogous to nuts distributed on cake-top. The inter-clump medium is non-stationary and clump-merger is likely triggered by the motion of this medium that drags the clumps floating in it. To examine if this was indeed the case, we have, in the lower-panel of Fig. 4, plotted rendered images of gas pressure in the region of merging clumps shown in the upper-panel of this same figure. The gas pressure for this plot is a sum of the thermal and the non-thermal contributions (by which we mean the kinematic pressure due to the velocity of gas particles, the magnitude of which has been discussed in \S 4.1b below). From these plots it is evident that regions of relatively low pressure marked in yellow-red shades join to form a single elongated structure. The density field is now segregated into the warm and cold phase. However, this observation does not support the hypothesis suggested by Koyama \& Inutsuka (2004), according to which clumps must have a higher pressure relative to the inter-clump medium so they could wade through this medium to make merger possible. The difficulty with this hypothesis is that clumps would disperse quickly if they were to have a pressure greater than that of the medium in which they reside; there is no evidence to suggest dispersal of clumps in these simulations.

Temperature distribution of molecular gas within the fragmented slab at two epochs has been shown in the rendered plots of Fig. 5. Cooler and relatively dense regions in the fragmented slab can be identified as those shaded red on this image. The rendered image on the left-hand panel of this figure shows the gas temperature within the fragmented slab at $t$ = 2.21 Myrs which in conjunction with the corresponding density plot in Fig. 3 can be used to estimate the length of the fastest growing mode, $L_{fast}$, defined by Eqn. (2) above. A simple calculation yields, $L_{fast}\ \sim$ 1.3 pc, which, as can be seen on the central-panel of the second row of Fig. 3, is indeed consistent with the length of fragments visible in this image. The picture on the right-hand panel of Fig. 5 shows the distribution of gas temperature at the time when calculations for this case were terminated ($t\sim$ 6.1 Myrs). The red areas on this image, and therefore colder regions in the fragmented slab, denote regions where  smaller cores  have probably begun forming via sub-fragmentation of larger clumps. The rarefied gas between clumps, which in the rest of this article will be described as the inter-clump medium, is an order of magnitude warmer than the gas in the dense-phase. These two phases, as expected, are in approximate pressure equilibrium; the average density and temperature of the inter-clump medium is respectively $\sim 2.5\times10^{3}$ cm$^{-3}$, and $\sim$120 K which corresponds to a pressure, $\sim$3$\times\ 10^{5}$ K cm$^{-3}$, is comparable to the pressure in the dense-phase at $\sim 10^{4}$ cm$^{-3}$ and $\sim$30 K. 

Figures 3 and 5 further highlight the importance of cooling in the process of gas-fragmentation within clouds, and therefore, in the overall process of core-formation.  Note that the pictures shown in Fig. 5 were prepared by distributing particles over a grid of 1000 cells in each dimension so that temperature has been averaged over pixels and  could therefore, easily blanket some of the coldest particles in the calculation. Since the nature of the dynamical processes within the slab remain unaltered despite changes in gas temperature, we have shown here plots only from case 2 listed in Table 1 above. Below we will present a mutual comparison of the dynamical processes that lead to fragmentation of the slab in each of the three realisations. This will be followed by a comparison of other diagnostic properties of gas in the post-fragmentation slab.

\begin{figure*}
  \vspace*{10pt}
   \includegraphics[angle=270,width=\textwidth]{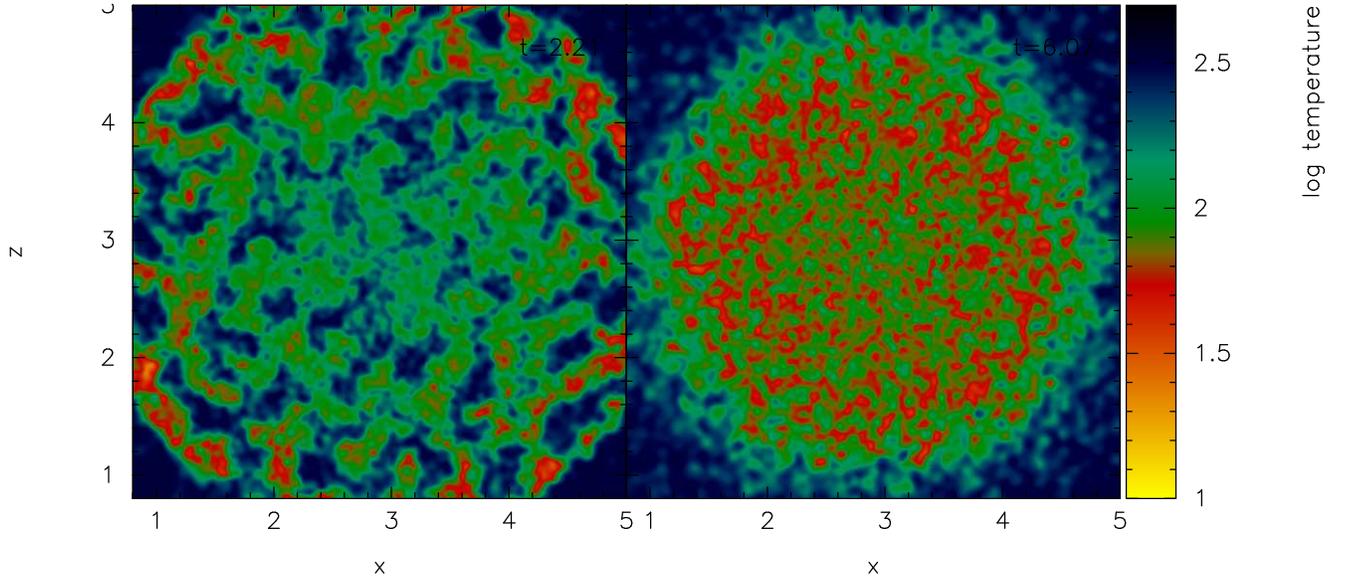}
  \caption{Rendered images of the distribution of gas temperature in the mid-plane of the slab in case 2 for two epochs($t$=2.21 Myrs and 6.07 Myrs for the plot respectively on the left and right-hand panel), has been shown here.}
\end{figure*}


\begin{figure}
  \vspace*{10pt}
  \mbox[\includegraphics[angle=270,width=0.5\textwidth]{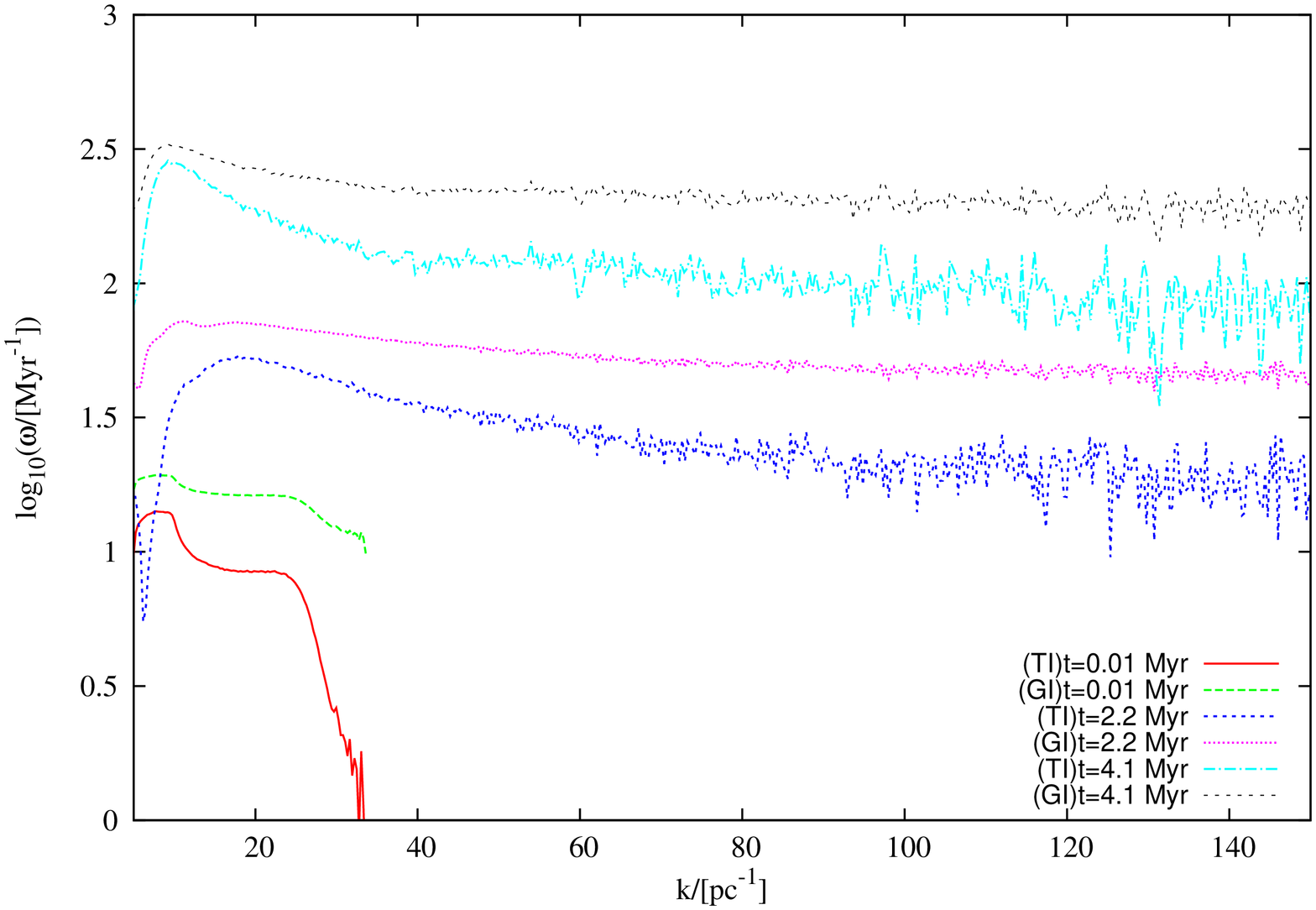}
        \includegraphics[angle=270,width=0.5\textwidth]{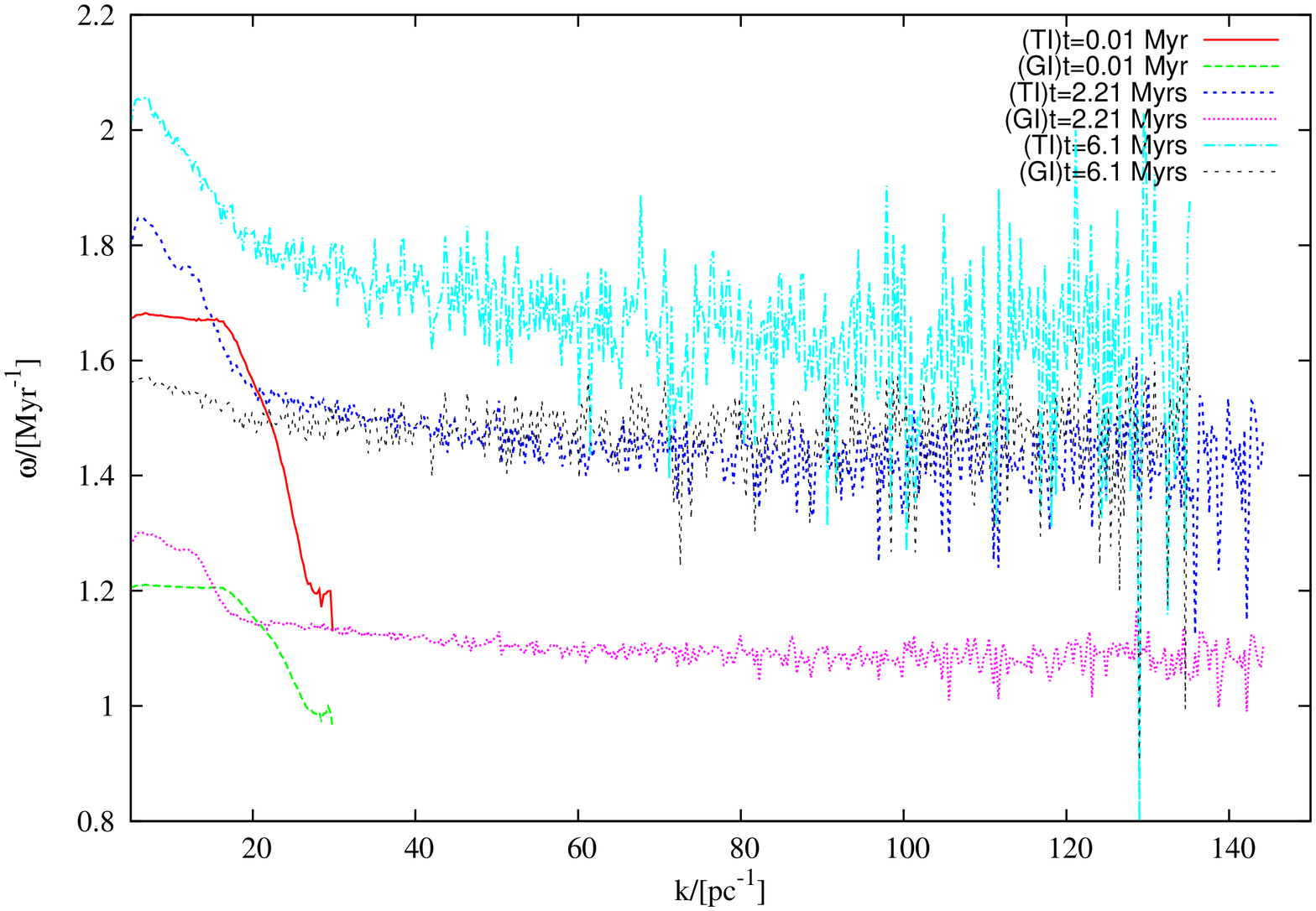}
        \includegraphics[angle=270,width=0.5\textwidth]{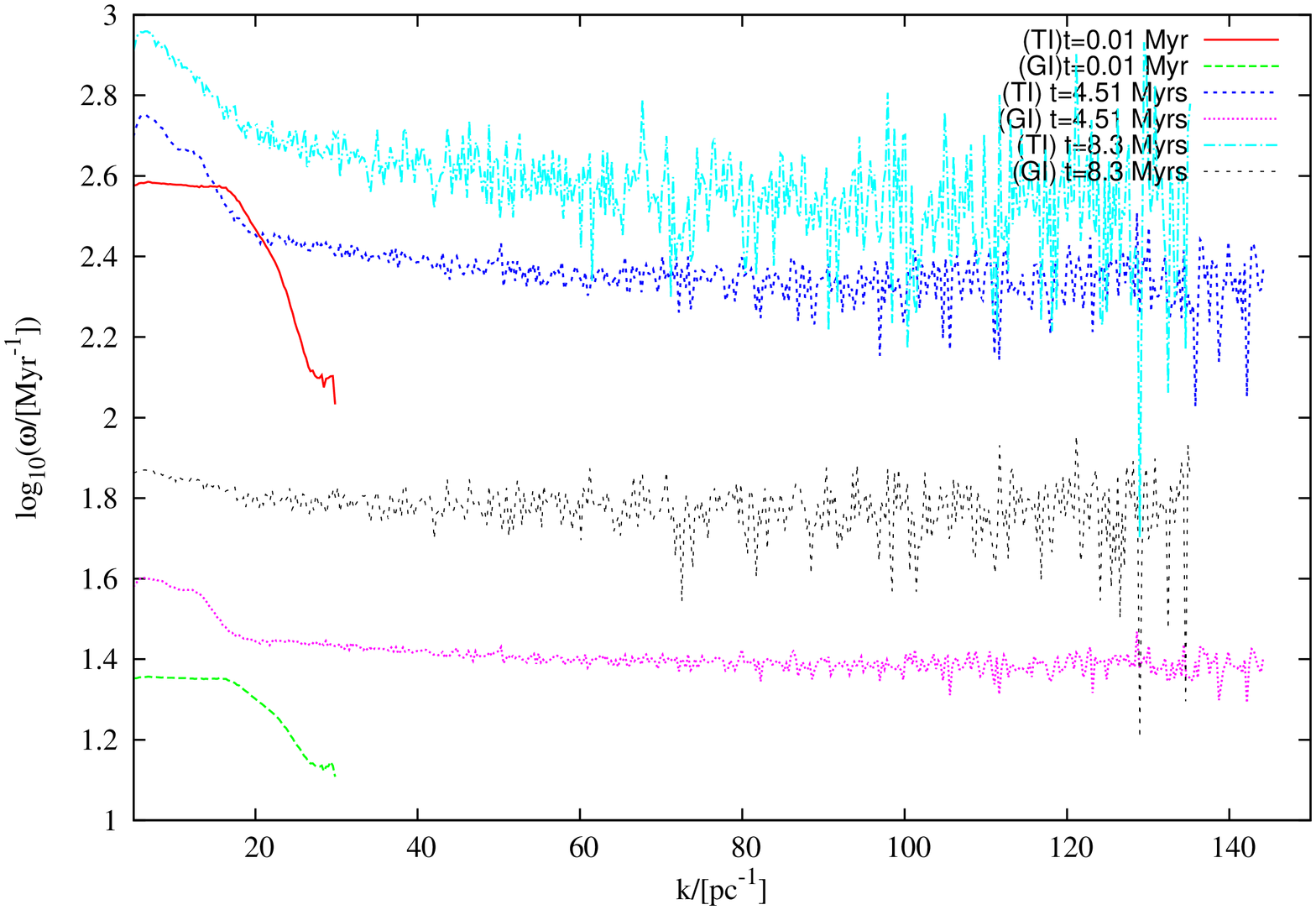}]
  \caption{Comparative plots of the growth rates of the gravitational and the thermal instability in cases 1, 2 and 3 have been shown on respectively, the top, middle and the lower panel.}
\end{figure}

\subsection{Dynamical instabilities within a cooling slab}
The slab evolves in the wake of an interplay primarily between two instabilities, viz. (\textbf{i}) the thermal instability ({\small TI}) and (\textbf{ii}) the gravitational instability ({\small GI}).
As is common-knowledge, a gas-body becomes susceptible to the {\small TI} when it is perturbed from its state of thermal equilibrium. In these simulations we achieve this by setting the initial gas-temperature above its equilibrium temperature($T_{eq}\sim$ 30 K; initial equilibrium pressure, $\frac{P_{eq}}{k_{B}}\sim$ 7500 K cm$^{-3}$), as dictated by the cooling-curve shown in Fig. 2.
The calculations steadily converge to a new equilibrium state that depends on gas density; we are essentially looking to minimise the quantity, $\mathcal{L}\equiv (n^{2}\Lambda(T) - n\Gamma)$, for gas number-density, $n$, and $\mathcal{L}\rightarrow 0$  close to equilibrium temperature. The absolute timescale over which the slab attains its thermal-equilibrium therefore does not really carry much physical significance and must be interpreted in a comparative sense only. Without the thermal conductivity we cannot define the Field length  in these calculations and so, the minimum length-scale for the {\small TI}, in these simulations, is supplanted by the {\small SPH} smoothing length. A volume of gas becomes susceptible to the {\small TI} if, 
\begin{equation}
\frac{\partial \mathcal{L}}{\partial T}\Big\vert_{p} < 0
\end{equation}
(Field 1965); $\mathcal{L} = n^{2}\Lambda - n\Gamma$. However, the gas may also be destabilised due to
\begin{equation}
\frac{\partial \mathcal{L}}{\partial T}\Big\vert_{\rho} < 0
\end{equation}
(Clarke \& Carswell 2007). Equivalently, while the Field criterion must be satisfied for the gas to become thermally unstable, it may also become unstable according to the latter condition. In the present work, the slab, initially held at uniform pressure becomes thermally unstable most likely by satisfying the Field criterion and then, during the course of its evolution, the next criterion is also likely satisfied as gas is segregated into the dense and the rarefied phase.

 Furthermore, if at least one side of the gas is also confined by ram-pressure, the slab-{\small ICM} interface demonstrates propensity towards growth of corrugations such as those associated with the thin-shell instability ({\small TSI}) (Vishniac 1983). In the present work though, thermal pressure confines the slab and the amplitude of perturbations on the slab-{\small ICM} interface are much smaller than the thickness of the slab itself. So in the rest of this article we will refer to the growth of the corrugations as the shell-instability, keeping in mind that the corrugations appear only on this interface. 

 During the course of evolution, as perturbations in the density field begin to grow gas within the slab gets segregated into the dense and rarefied phase. Each subsequently acquired state of thermal equilibrium is impermanent as density and pressure adjustments happen quickly and the gas is no longer in gravo-thermal equilibrium. The implications of this observed fragmentation for the evolution of molecular clouds will be discussed in \S 5.1. The typical fragmentation length-scale of this cooling gas is given by Eqn. (2).  Of the two instabilities,  the {\small TI} grows more rapidly in comparison to the {\small GI}
in realisations 2 and 3 whereas the latter dominates the former in the first realisation. The image on the central-panel of Fig.3, for instance, shows that the mid-plane of the slab becomes flocculated as it cools simultaneously. 

In an earlier contribution we have demonstrated that growth of the shell-instability induces a velocity field in the underlying gas and
 have explored the non-linear excursion of this instability(Anathpindika 2009). In this latter case, the amplitude of perturbations become comparable to the slab thickness which is not the case here. In this work, however, we are interested in examining the growth of the {\small GI} in a cooling slab. For  although the {\small TI} initially generates structure in the density field, the {\small GI} sets in as  gas continues to cool and induces sub-fragmentation once the Jeans condition is satisfied. It must be remembered though, the fragments had not become isothermal when simulations were terminated and so the thermal Jeans length can only serve as a rough estimate for the typical size of a core (or a clump, in general). The thermal Jeans length, as is well-known, is defined as,
\begin{equation}
\lambda_{Jeans} = \Big(\frac{\pi a^{2}}{G\mu\bar{n}}\Big)^{1/2}
\end{equation}
(Jeans 1928), which at $\bar{n}\sim$ 2$\times$10$^{4}$ cm$^{-3}$, our choice of density threshold to detect clumps,  and temperature, $T_{gas}\sim 10$ K, becomes, $\lambda_{Jeans}\sim$ 0.2 pc, which is significantly larger than the extent of the smallest resolvable spatial region in these simulations and satisfies the Truelove criterion for resolution so that artificial fragmentation is unlikely. Although we have demonstrated here the process of clump-formation using only a relatively simple cooling function, a more realistic treatment of the process in which molecular gas attains thermal equilibrium would need the appropriate molecular chemistry (see e.g. Glover \emph{et al.} 2010). We leave this for a future work.  

\begin{figure}
  \vspace*{10pt}
       \includegraphics[angle=270,width=0.52\textwidth]{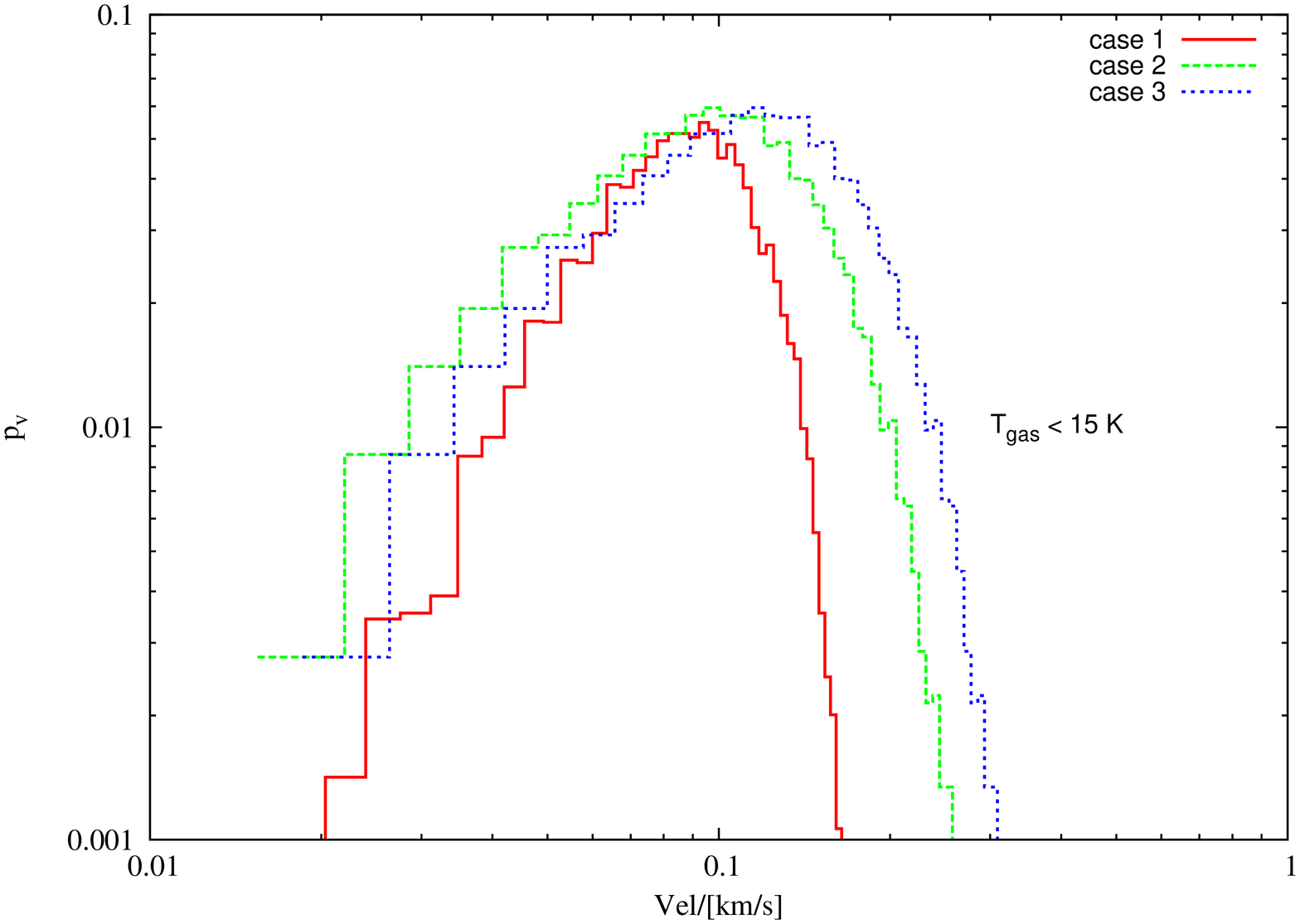}
        \includegraphics[angle=270,width=0.52\textwidth]{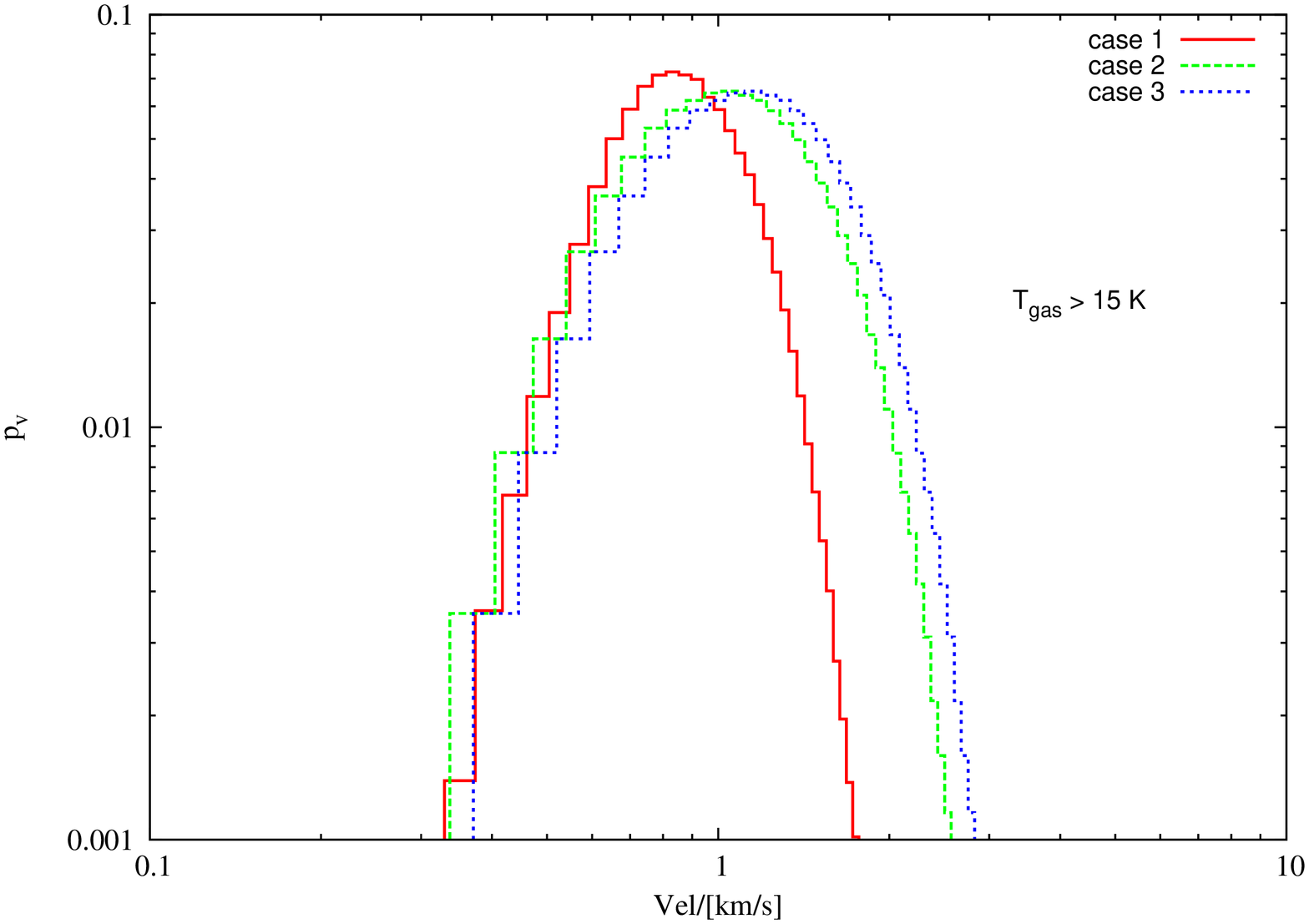}
 \caption{Shown here are the plots of the local velocity distribution for gas in the cold and warm phase within the fragmented slab at the time of terminating calculations for each of the three realisations.}
\end{figure}

\begin{figure}
 \vspace*{10pt}
        \includegraphics[angle=270,width=0.52\textwidth]{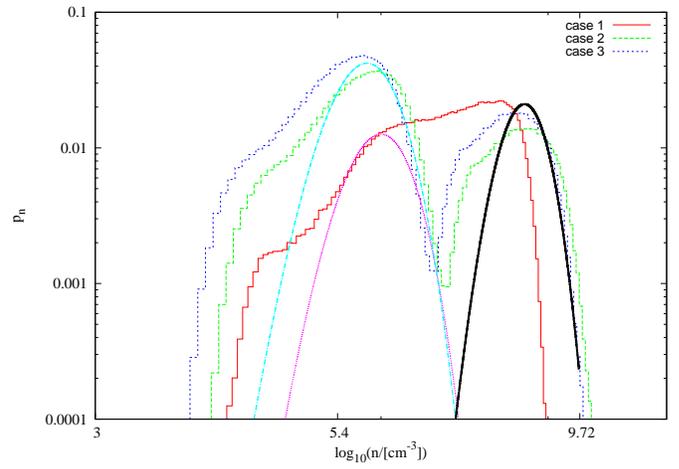}
  \caption{The density probability distribution for gas within the fragmented slab in each of the three test cases has been shown here. As for the velocity PDFs shown in Fig. 7, this plot was also made at the time of terminating calculations. Apart from the difference in timescale on which these PDFs take shape, the distributions for cases 2 and 3 are bi-modal while that for case appears to be developing a power-law tail at higher densities.}
\end{figure}

\begin{figure}
 \vspace*{10pt}
        \includegraphics[angle=270,width=0.52\textwidth]{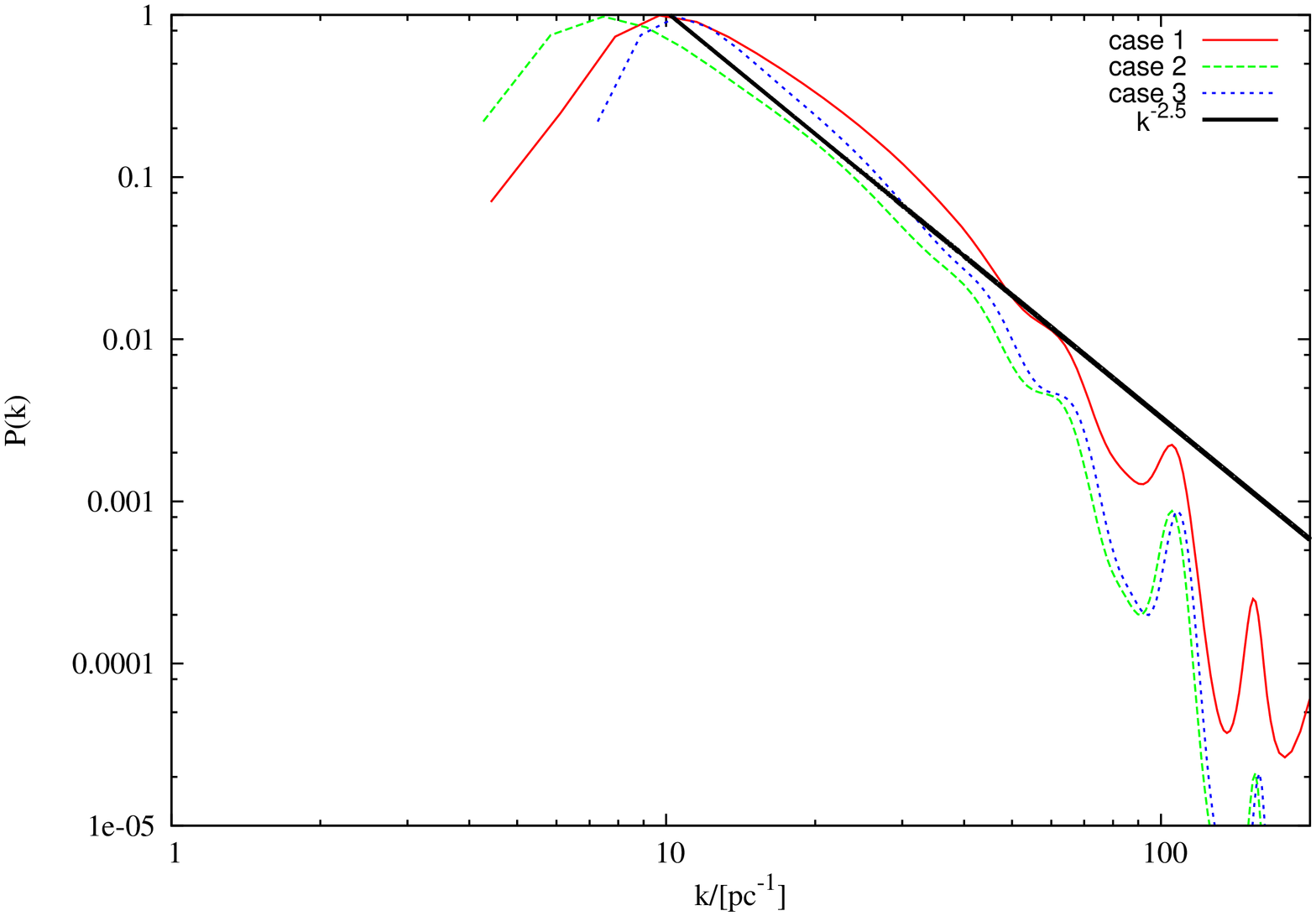}
  \caption{The power-spectrum for the three cases has been shown here and was derived at the same epoch as that for plots in Figs. 7 and 8 above. Individual spectra at large wave-numbers \textbf{are steeper} than the Kolmogorov power-spectrum; see text for description. }
\end{figure}

At this point a comparison of rates at which the {\small TI} and the {\small GI} grow, will prove useful in unravelling the fragmentation process. It will help us understand which of the two instabilities are responsible for clump-formation in a {\small MC}.
During fragmentation it is likely, cooling at a given pressure and density, both have a destabilising effect on gas within the slab. In this case the {\small TI} grows approximately on the thermal timescale as perturbations need not wait for a pressure equilibrium to become unstable. The thermal timescale - 
\begin{equation} 
\tau_{TI}\ \equiv\ \frac{E_{therm}(T)}{(n\Lambda(T) - \Gamma)},
\end{equation}
so that its growth-rate,
\begin{equation}
\omega_{TI}\ =\ \frac{1}{\tau_{TI}}.
\end{equation}
The growth-rate, $\omega_{TI}$, given by Eqn. (8) above was calculated by distributing the {\small SPH} particles on a 256$^{3}$ grid in the $k$-space. The average density weighted temperature for each cell on this grid was then calculated and followed by the evaluation of the cooling function, $(n\Lambda(T) - \Gamma)$, for that cell.
The expression for the growth-rate of the {\small GI} on the surface of the slab, on the other hand, from the usual Jeans-analysis, is 
\begin{equation}
\omega_{J}\ \equiv\ (4\pi G\mu\bar{n})^{1/2}
\end{equation}
(Jeans 1928), which is also similar to that suggested by Elmegreen \& Elmegreen(1978) and Ledoux (1951), though the former  suggest growth-rate higher by a factor of 2 than that given by Eqn. (9), and therefore, is unlikely to alter conclusions drawn here.  

\begin{figure}
  \vspace*{10pt}
  \mbox[\includegraphics[angle=270,width=0.5\textwidth]{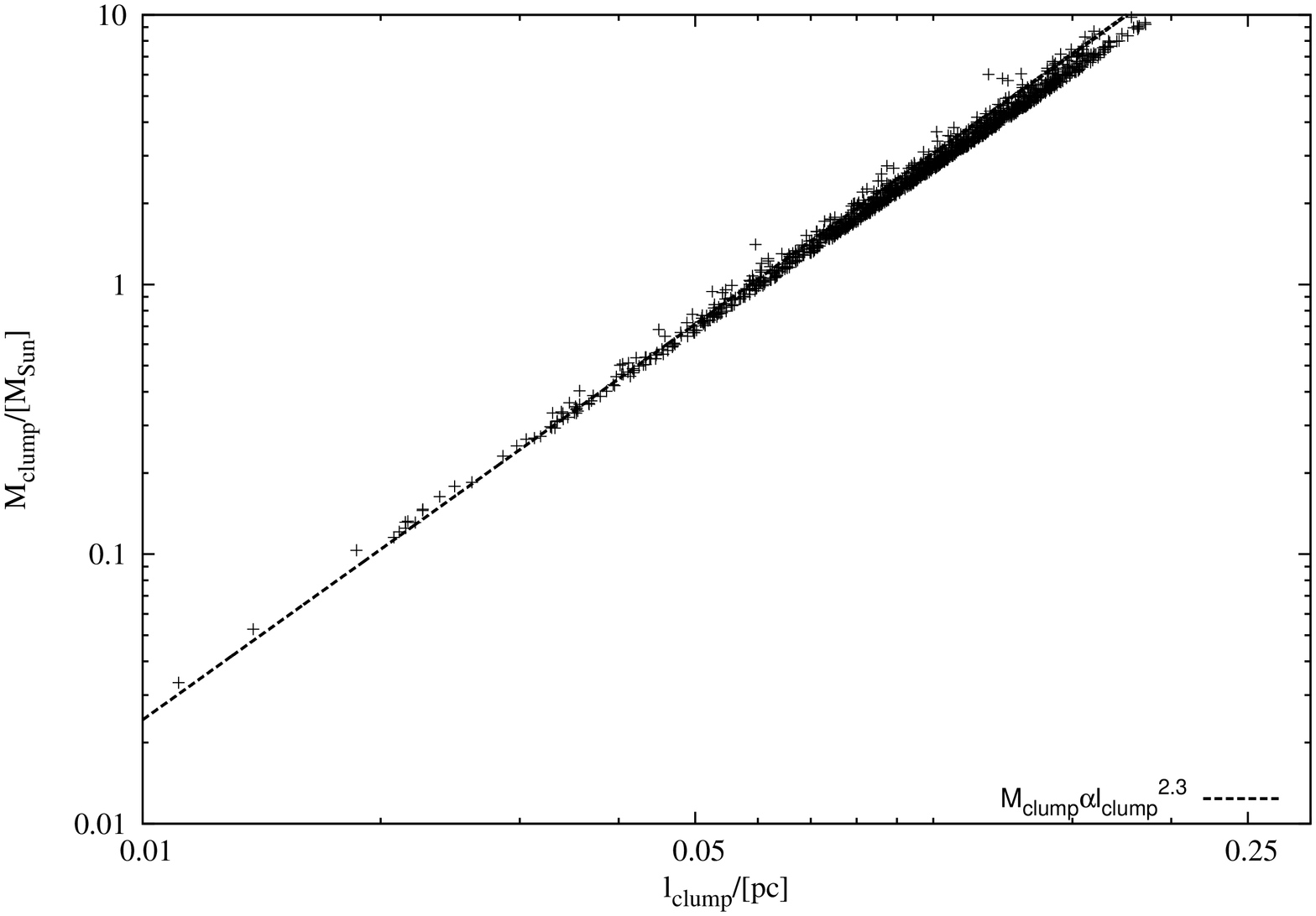}
        \includegraphics[angle=270,width=0.5\textwidth]{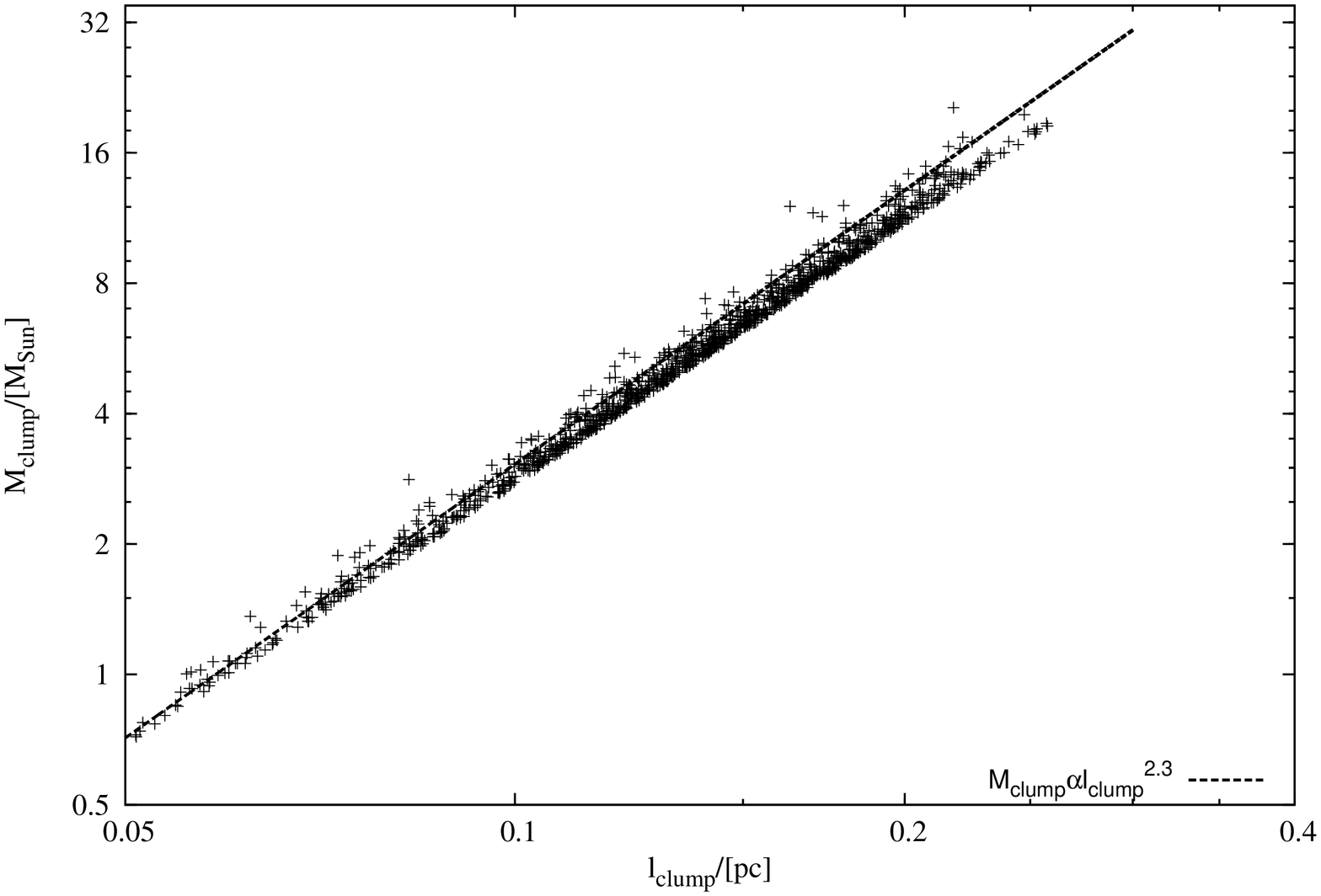}
        \includegraphics[angle=270,width=0.5\textwidth]{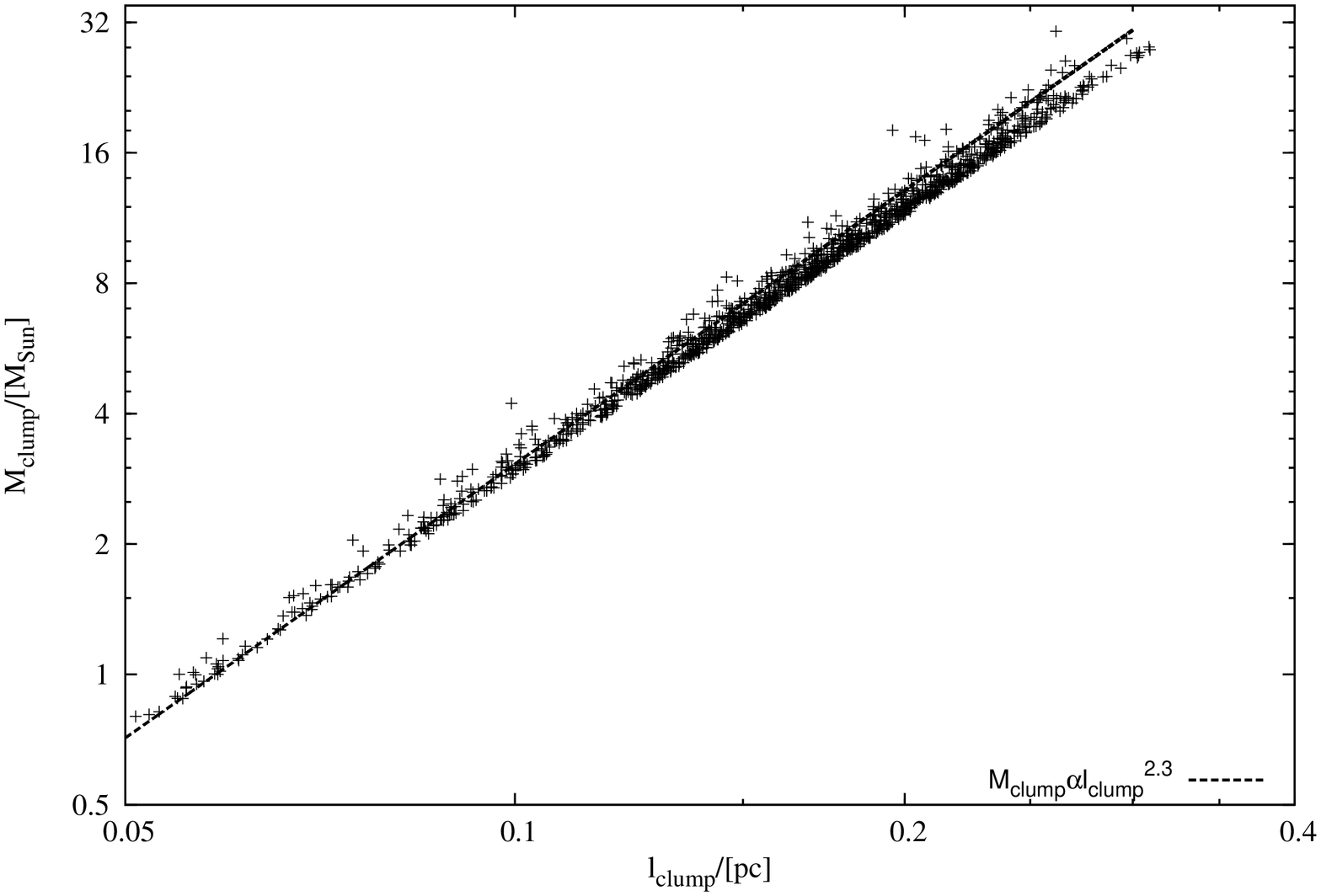}]
  \caption{The upper, middle and the lower panel shows the distribution of mass against the size of cores detected in each of the three realisations. The correlation is tight and well approximated by the $M_{core}\propto L_{core}^{3}$.}
\end{figure}

Plotted on the upper, middle and the lower panel of Fig. 6 are respectively, the measured growth rates  for the {\small TI} and the {\small GI} for cases 1, 2 and 3. It is clear that in each of the three test cases the slab fragments via a confluence of the {\small TI} and the {\small GI}. A common feature visible in these plots for cases 2 and 3 is the subservience of the {\small GI} relative to the {\small TI} suggesting, the fragmentation is largely driven by the {\small TI}, the growth of which is the strongest in case 3. On the contrary, the {\small GI} grows more strongly as compared to the {\small TI} in case 1, where the initial distribution of gas in the slab was gravitationally super-critical (See Col. 12 of Table 1). Realisation 3 is the one that has the highest initial gas temperature and therefore departed strongly from thermal equilibrium (equilibrium temperature, $T_{eq}\ \sim$ 30 K). The {\small TI}-induced fragmentation is hierarchical as the slab first fragments on large spatial scales and then trickles down to smaller spatial scales, i.e. higher wave-numbers. Growth of either instability in each realisation begins on the length-scale, $L_{fast}$, defined by Eqn. (2) and can be seen from the plots shown in Fig. 6. 

Fall in the growth-rate of the {\small TI} at a length-scale indicates a lower rate of cooling at that length-scale which means gas there is closer to acquiring thermal equilibrium. The reduction in the growth-rate of the {\small TI}, as is visible from these plots is impermanent and picks-up as the slab continues to fragment. The situation in the fragmenting slab is dynamic so that pressure adjustments happen quickly. The relatively large fragments break up further as they continue to cool, a process that is simultaneously associated with the steadily rising strength of gravity. This is evident from the observed rise in the growth-rate of the {\small GI} with passage of time. Also, it appears, the {\small TI}-induced fragmentation was not complete at the time of terminating these calculations. However, in all the test cases its growth does appear to be slowing at higher wave-numbers (smaller spatial-scales ), as it should when fragments are approaching their equilibrium temperature. This slowing is more prominently visible in the plots for Case 1 where the slow-down is visible at all wave-numbers. As noted earlier, we could not follow the later stages when these smaller fragments would likely begin to collapse. 
In the following section, we will separately examine some physical properties of the gas and the clumps detected in these simulations.

\section{Some diagnostic properties of gas}
As we are concerned with the behaviour of gas in  the prestellar phase, it is crucial to examine the distribution of its other physical properties.  In this regard, we call attention to the distribution of gas velocity and density before examining the properties of cores detected in each realisation.

(1) \emph{The velocity probability distribution function ({\small PDF})}
Of significant importance to the process of fragmentation is the behaviour of the relatively cool gas, typically found at a few ten Kelvin. We therefore describe as cold, gas cooler than 15 K, and warm, that is at temperature higher than 15 K. This threshold of gas temperature is suitable to study the kinematic properties of potential star-forming gas. Histograms showing the distribution of gas velocity on either side of the temperature threshold have been plotted on the upper and the lower panel of Fig. 7 for each of the three realisations. We note that apart from the variation in the magnitude of velocity, the respective histograms show little mutual variation. The magnitude of gas velocity increases successively with increasing initial gas temperature in the three realisations. Also, in each of these cases the histogram for cold gas peters off steeply at the high velocity end. This suggests, velocity field within a potential star-forming clump is likely to have a magnitude lower than a certain turnover value, perhaps determined by the prevailing physical conditions.
It is, however, difficult to draw any specific inferences from these {\small PDF}s, or in general, from similar ones derived for {\small {\small MC}s}. At best they are indicative of the relative strength in the velocity bins.  Observationally deduced velocity {\small PDF}s for typical star-forming clouds are known to have a variety of shapes. For instance,  similar looking {\small PDFs} have been derived for {\small {\small MC}s} such as the Orion, Monoceres and a few others by Miesch \& Scalo (1995).  Interestingly, velocity magnitudes seen here are also comparable to those reported for gas within these clouds (e.g. Curtis \emph{et al.} 2010; Falgarone \emph{et al.} 2009; Hatchell \emph{et al.} 2005 and references there-in). 
 
(2)\emph{The density probability distribution function}
Shown in Fig.8 are density {\small PDF}s for each of the three realisations, plotted at the time calculations were terminated in each case. Although similar looking, the {\small PDF}s for cases 1 and 2 evolve on a timescale that is significantly different although it is comparable for cases 2 and 3. This is because growth of the {\small TI} is much more sluggish in case 1 as compared to that in the remaining two realisations as was demonstrated in Fig. 6. The {\small PDF}s for the latter two realisations viz., 2 and 3, are bi-modal, a combination of two log-normal distributions, which is consistent with the behaviour of a 2-phase medium. The log-normal fit to these {\small PDFs} was obtained using the following expression,
\begin{equation}
p(\ln\ n) = \frac{1}{(2\pi \sigma^{2})^{1/2}}\exp\Big[-\frac{(\ln\ n - \ln\bar{n})^{2}}{2\sigma^{2}}\Big]
\end{equation}
(e.g. Padoan \emph{et al.} 1997; Federrath \emph{et al.} 2008; Audit \& Hennebelle 2010; Federrath \emph{et al.} 2010 and Gazol \& Kim 2013); $\sigma$, being the full-width at half-maximum.  The {\small PDF} for case 1, on the other hand, appears to be developing a power-law tail at higher densities. However, this latter stage when fragments could begin to collapse was not followed in these simulations (see e.g., Klessen \& Burkert 2000; Dib \& Burkert 2005; V{\' a}zquez-Semadeni \emph{et al.} 2008; Kritsuk \emph{et al.} 2011; Ballesteros-Paredes \emph{et al.} 2011 and Federrath \& Klessen 2013).

{\small PDF}s have now been derived for a number of Galactic {\small MC}s. For instance, Andr{\' e} \emph{et al.}(2010) have derived one for the Aquila star-forming cloud while Kainulainen \emph{et al.}(2009), for nearly two-dozen other Galactic {\small {\small MC}s} (see also Schneider \emph{et al.} 2012, 2013; and Kainulainen \emph{et al.} 2014). The density {\small PDF} for the Aquila {\small MC} has a distinct power-law tail at its high-density end. Similarly, other star-forming clouds such the Taurus and Lupus I also demonstrate a tail at the high-density end. On the contrary, relatively quiescent clouds such as Lupus V and Coalsack have approximately log-normal density {\small PDFs} (Kainulainen \emph{et al.} 2009). In general, however, the density {\small PDF} shows  considerable variation across {\small MC}s.

(3) \emph{Velocity power-spectrum}
Growth of the dynamical instabilities discussed above induces a velocity field within slab-layers.
 The power-spectrum is a vital diagnostic of energy distribution  across the spatial scales of a given velocity field (Federrath 2013). Observationally spectral indices have been determined for molecular clouds using emissions from the CO molecule and its isotopologues. Spectral indices in these cases have usually been suggested to lie between -2 and -4; see review by Hennebelle \& Falgarone (2012).  Shown in Fig. 9 is the spectrum for the three test cases generated at the time of termination of calculations in these respective realisations and evidently has two peaks. A double-peaked spectrum is often reported for {\small MC}s (e.g. Schneider \emph{et al.} 2011), and indicates the characteristic length-scale of dense structure in a \small{MC}. The spectra derived for realisations discussed in this work peak at $\sim$1 pc, the typical length of filaments and $\sim$0.2 pc, the characteristic size of clumps detected in the fragmented slab. However, unlike the spectra for typical star-forming clouds, the spectra derived here peters-off at small spatial-scales as protostellar feedback has not been included in these simulations. The spectrum was calculated by first interpolating the {\small SPH} particles on to a grid  having resolution 256$^{3}$ in the $k$-space. For this we followed the procedure described by Price (2007) while developing the publicly available visualisation algorithm {\small SPLASH}. Then the velocity vector field in the $k$-space was generated using the relation
\begin{equation}
\mathbf{v}(\mathbf{k}) = \frac{1}{(2\pi)^{3}}\int_{V}\mathbf{v}(\mathbf{r})\exp(-i\mathbf{k\cdot r})\ d\mathbf{r}.
\end{equation}
The corresponding power, $P(k)$, was then calculated according to
\begin{equation}
P(k) = <\mathbf{v(k)}\cdot \mathbf{v}^{\ast}(\mathbf{k})>.
\end{equation}

At the epoch when the spectrum was made, turbulence between wave-
numbers (1,100) has not decayed yet. There is, however, no evidence of a significant decrease in power at smaller wave-numbers, i.e. larger spatial scales. Consequently, relatively large fragments are still probably sub-fragmenting at this epoch. On smaller spatial scales, or higher wave-numbers though, the decrease in the spectrum shows that {\small TI} induced fragmentation has probably terminated. The resulting clumps on small spatial scales are thermally supported, although most at this stage could be unbound at this stage. The spectrum is significantly steeper than the Kolmogorov spectrum and can be reasonably well approximated by a power-law of the type, $P(k)\propto\ k^{-2.5}$. Thus, gas motions within the slab induced by pressure-confinement and the confluence of various instabilities described above, appear to be compressive in nature and interestingly, is similar to the spectra derived for simulations where isothermal supersonic turbulence is injected. In these later cases the power-spectrum is usually observed to have a slope slightly in excess of -2 (e.g. Padoan \emph{et al.}1997;  Kim \& Ryu 2005;  Price \& Federrath 2010; Federrath 2013). 

The other striking feature of these spectra is that each of them exhibits secondary peaks at higher wavenumbers, $k\gtrsim$ 100. As argued by Kitsionas \emph{et al.}(2009); Federrath \emph{et al.}(2010; 2011) and Price \& Federrath (2010), this feature and the somewhat excessive steepness of the spectra are both apparently the prognosis associated with inadequate numerical resolution, though it does not call into question the conclusion that dynamic instabilities in the slab generate compressive motions within its layers. Kitsionas \emph{et al.}(2009) argue that a minimum of 32 grid cells (i.e. 16 {\small SPH} smoothing kernels), are necessary to eliminate numerical artefacts that could possibly be introduced by dissipation mechanisms employed in a hydrodynamic code and more crucially, achieve numerical convergence in calculations of statistical quantities such as the power-spectrum associated with the underlying turbulent velocity field. The critical wave-number below which numerical convergence can be obtained is, $k_{crit}\le \Big(\frac{N=\sqrt[3] N_{gas}}{32}\Big)$, which for the current numerical exercise is $\sim$3. In other words, the turbulent power-spectrum derived here is not resolved on the typical length-scale of fragmentation observed in these realisations. In order to be able to resolve the spectrum on the length-scale of the fastest growing mode defined by Eqn. (2) above, we will have to increase the number of particles by at least 1 order of magnitude above that used in this work, and by up to 3 orders of magnitude to resolve it on the length-scale of a typical core defined by Eqn.(6) above. We can therefore neither associate much significance to the observed slope of the spectra derived here, nor associate a causal relationship between the observed slope and the gas motions generated by the dynamic instabilities attending the slab. The secondary peaks in the power-spectra visible at higher wavenumbers are likely due to the accumulation of energy in these wave-number bins as density builds up on the corresponding length-scales.This problem would also be alleviated with an enhanced resolution as the observed accumulation of power in specific wave-vector bins will be more evenly distributed over smaller bins.

(4)\emph{Mass-size correlation}
Shown in  the top, middle and the lower panel of Fig. 10 is the correlation between the mass and size of the fragments detected in realisations 1, 2 and 3, respectively. The fit,  $M_{clump}\propto L_{clump}^{p}$, with $p=2.3$, for these test cases is remarkably tight. It is also well above the $M_{clump}\propto L_{clump}^{2}$ relation that signifies constant column density for the detected sample of clumps and their bounded nature. Interestingly, isothermal simulations by Federrath \emph{et al.} (2009), for instance, yielded a similar correlation, $M_{clump}\propto L_{clump}^{2.3}$, for clumps detected in simulations of compressionally driven turbulence. On the other hand, for simulations of solenoidally driven turbulence they deduced a much steeper correlation, $M_{clump}\propto L_{clump}^{2.6}$. 
Elmegreen \& Falgarone (1996) have reported a similarly steep correlation, $M_{clump}\propto L_{clump}^{2.35}$, for CO clumps.
Also, cores have been found in the Serpens and Ophiuchus {\small MC}s with a similarly steep slope such as those reported here (e.g. Enoch \emph{et al.} 2008). Similar slopes have also been reported for CO clumps by a few other authors, e.g. Heithausen \emph{et al.}(1998), Kramer \emph{et al.}(1998) and Falgarone \emph{et al.}(2004). Other recent observations of star-forming clouds by for instance, Roman-Duval \emph{et al.} (2010) and Donovan-Meyer \emph{et al.} (2013) also suggest a similar correlation. This suggests, star-forming clouds are likely governed by supersonic, compressive turbulence that generates dense sheets of gas.

\begin{figure}
  \vspace*{10pt}
        \includegraphics[angle=270,width=0.5\textwidth]{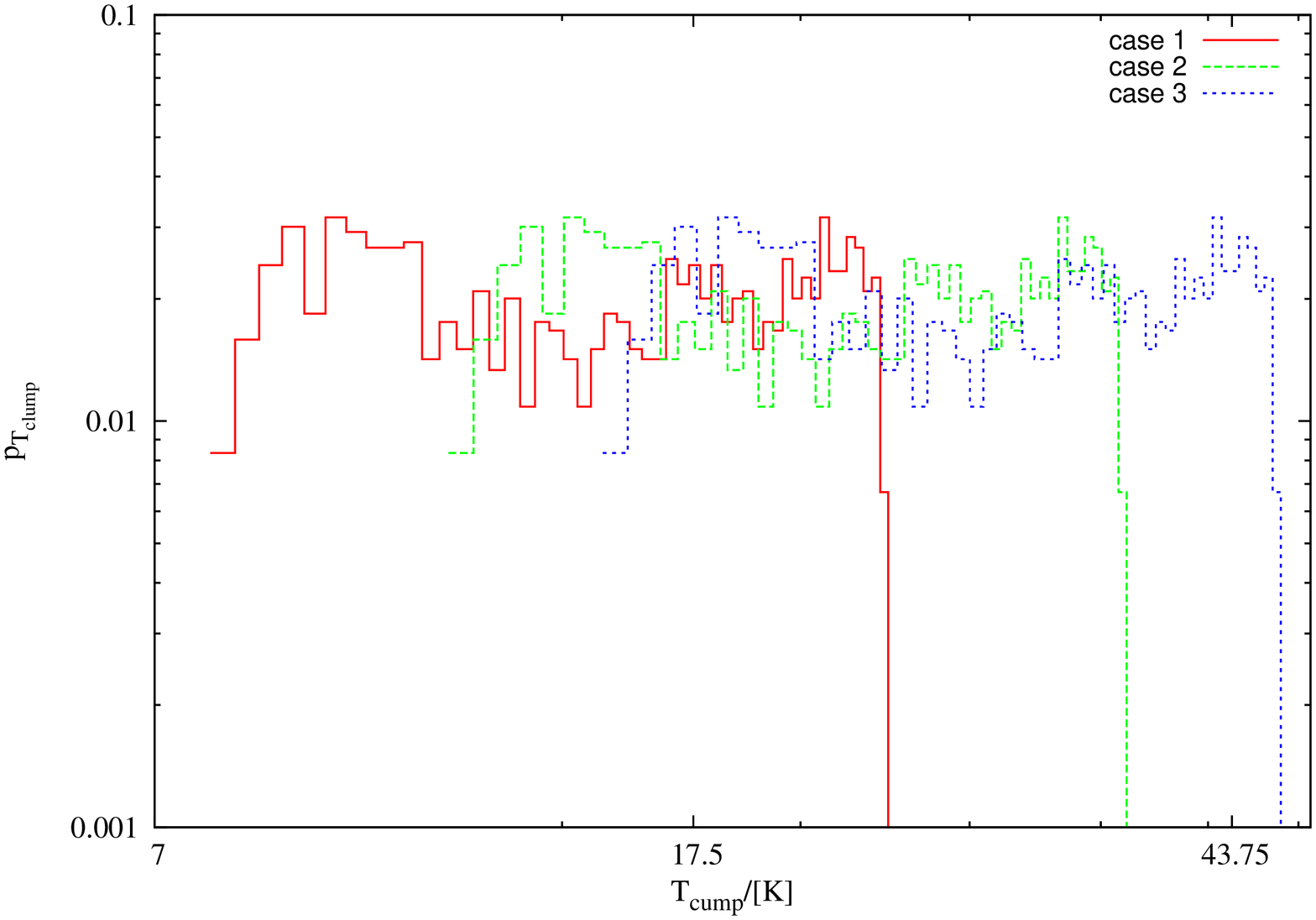}
  \caption{Shown here is the distribution of gas-temperature in clumps detected in cases 1, 2 and 3.}
\end{figure}

\subsection{Diagnostic properties of clumps}
We have seen in the above sub-section that despite the similarity in the general evolutionary sequence of the slab for different choices of the confining pressure, $p_{ext}$, the timescale of evolution and the magnitude of physical entities such as gas velocity and density vary across the three realisations.  Consequently, differences in the properties of clumps condensing out of this gas across the three test cases are not altogether unexpected. Below we will examine some physical properties of prestellar cores detected in each realisation. \\ \\
(a) \emph{Gas temperature in clumps} 
The distribution of gas temperature in the clumps detected in each of the three realisations has been shown in Fig. 11. The nature of temperature distribution is similar for each of the three realisations, however, the coldest clumps were found in realisation 1, followed by those that are successively warmer, in the remaining two realisations. Also, the dual-peaked temperature distributions suggest, one set of clumps that is relatively colder than the other. The implication of which is, this latter set of clumps is likely to sub-fragment as they continue to cool. In what follows we will investigate if other physical properties such as the mass and size of these clumps also depend on their temperature. 
 Observational surveys of typical star-forming clouds such as those by Jijina \emph{et al.}(1997) and Enoch \emph{et al.}(2008), have deduced temperatures in the range of 7 K - 40 K for prestellar cores and gas clumps. Apparently, this deduced temperature distribution peaks at about 12 K, though cores with even cooler molecular gas have also been reported frequently. We get a similar distribution for clumps detected in our simulations with the coldest clumps being detected in realisation 1. \\ \\
\begin{figure}
  \vspace*{10pt}
        \includegraphics[angle=270,width=0.5\textwidth]{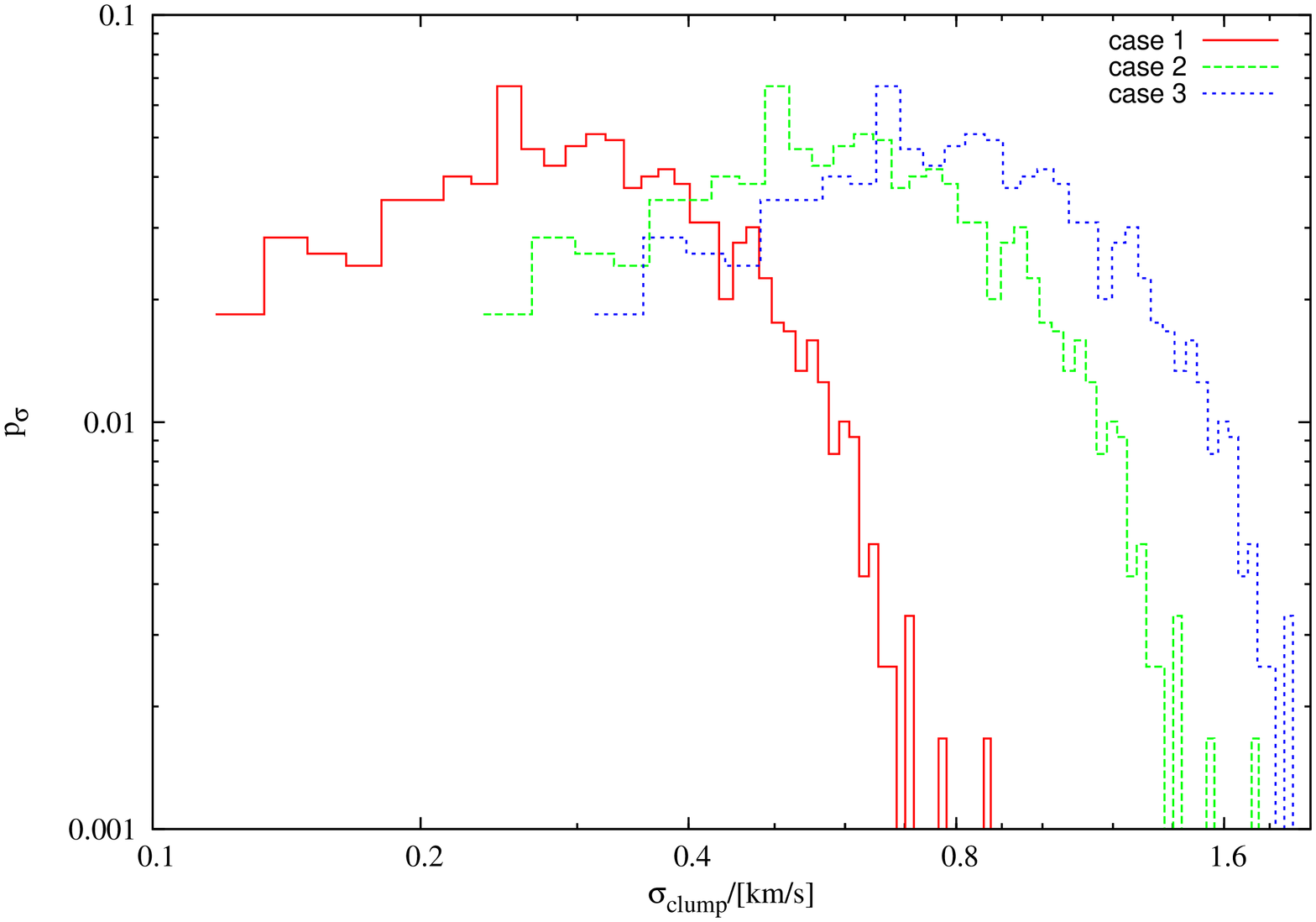}
  \caption{As in Fig. 11, the distribution of velocity dispersion for gas in the clumps detected in our simulations has been shown in these plots. }
\end{figure}

(b) \emph{Velocity field within clumps}
The histogram showing the distribution of gas velocity dispersion, $V$, within the clumps detected in the three realisations has been shown on the top right-hand panel of Fig. 12. Although the nature of the {\small PDF} is similar for the three test cases, the magnitude of velocity dispersion increases successively from case 1 through to 3. This is probably due to the stronger growth of the the thermal instability ({\small TI}) that triggers early fragmentation of the slab. In particular, the growth of this instability is stronger in cases 2 and 3 in comparison to that in case 1. That the {\small TI} indeed, grows strongly in these two realisations has been demonstrated above. Also, a stronger confining pressure also assists stronger growth of corrugations on the slab-{\small ICM} interface, associated with the shell-instability. Growth of this latter instability on the slab-{\small ICM} interface causes the interface to buckle as density perturbations amplify rapidly. This amplification is associated with momentum transfer between local peaks and troughs in the density field. A stronger confining pressure assists momentum transfer between density perturbations and therefore, their amplification. The combined effect of these instabilities probably holds the key to the observed velocity field in the detected clumps. Interestingly, the  magnitude of this velocity field is consistent  with that found within clumps and cores in typical star-forming regions; see e.g. Enoch \emph{et al.}(2008) and other references there-in. \\ \\
(c) \emph{Clump-size} The calculation of clump-size was discussed earlier in \S2.1 where the algorithm to detect clumps was described. The resulting distribution of clump-size for those detected in the respective test cases has been shown in Fig. 13. The detected clumps have sizes comparable to those of potential star-forming clumps and reported in previously cited literature.
Evidently, the sub-fragmentation of larger fragments into smaller cores must be still underway at the time of terminating these calculations. The sub-fragmentation of gas filaments, as can be seen in the rendered density plots of Fig. 3, occurs on the scale of the local Jeans length. For instance, the temperature distribution of clumps detected in case 2 peaks at $\sim$15 K, which at the clump-detection threshold of $\sim$10$^{4}$ cm$^{-3}$, corresponds $\lambda_{Jeans}\sim$ 0.2 pc, which is indeed where the distribution of clump-sizes for case 2 peaks. This is also true for the other two realisations.

\begin{figure}
  \vspace*{10pt}
        \includegraphics[angle=270,width=0.5\textwidth]{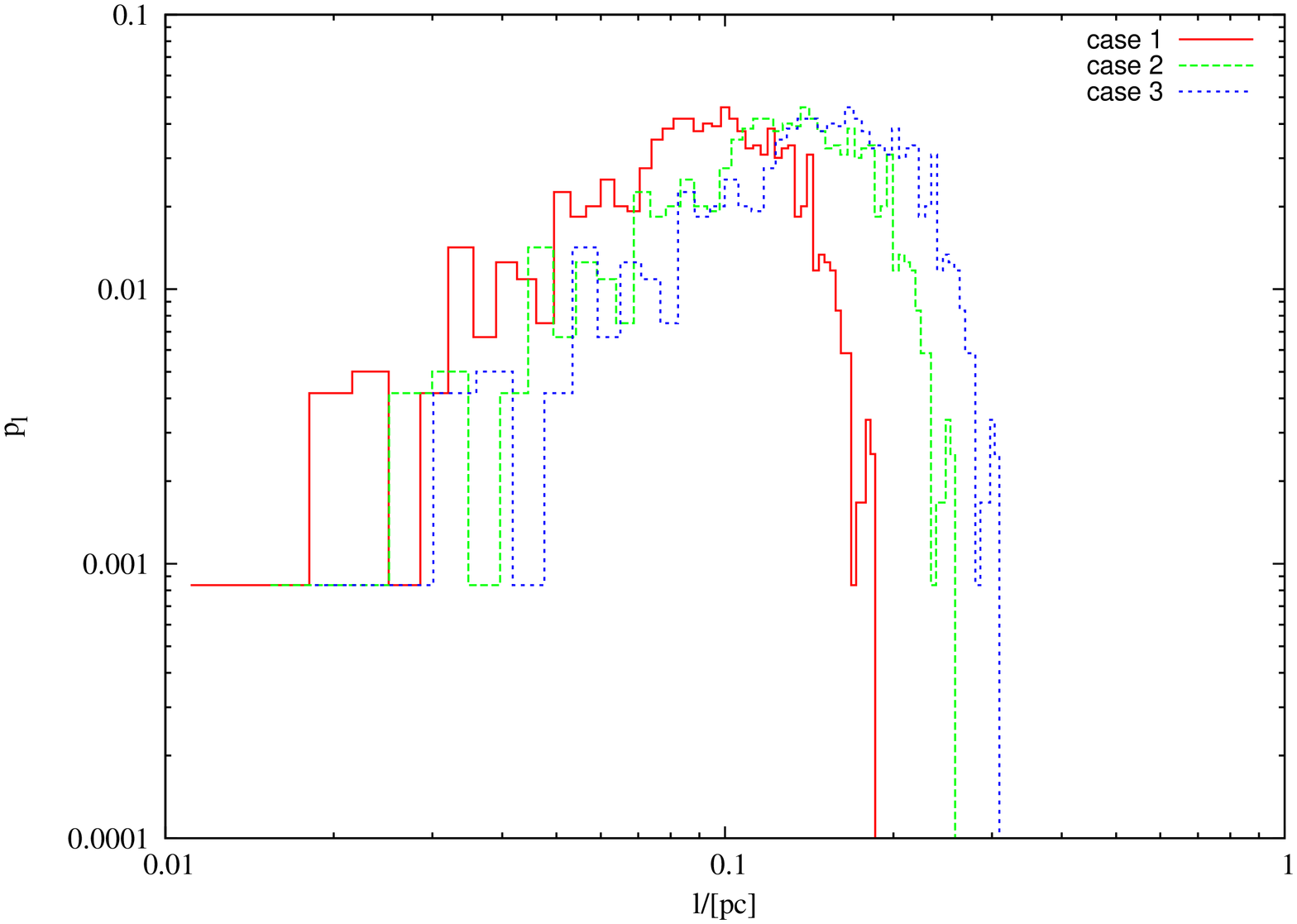}
  \caption{The distribution of sizes for clumps detected in cases 1, 2 and 3 has been shown in these plots.}
\end{figure}

\begin{figure}
  \vspace*{10pt}
        \includegraphics[angle=270,width=0.5\textwidth]{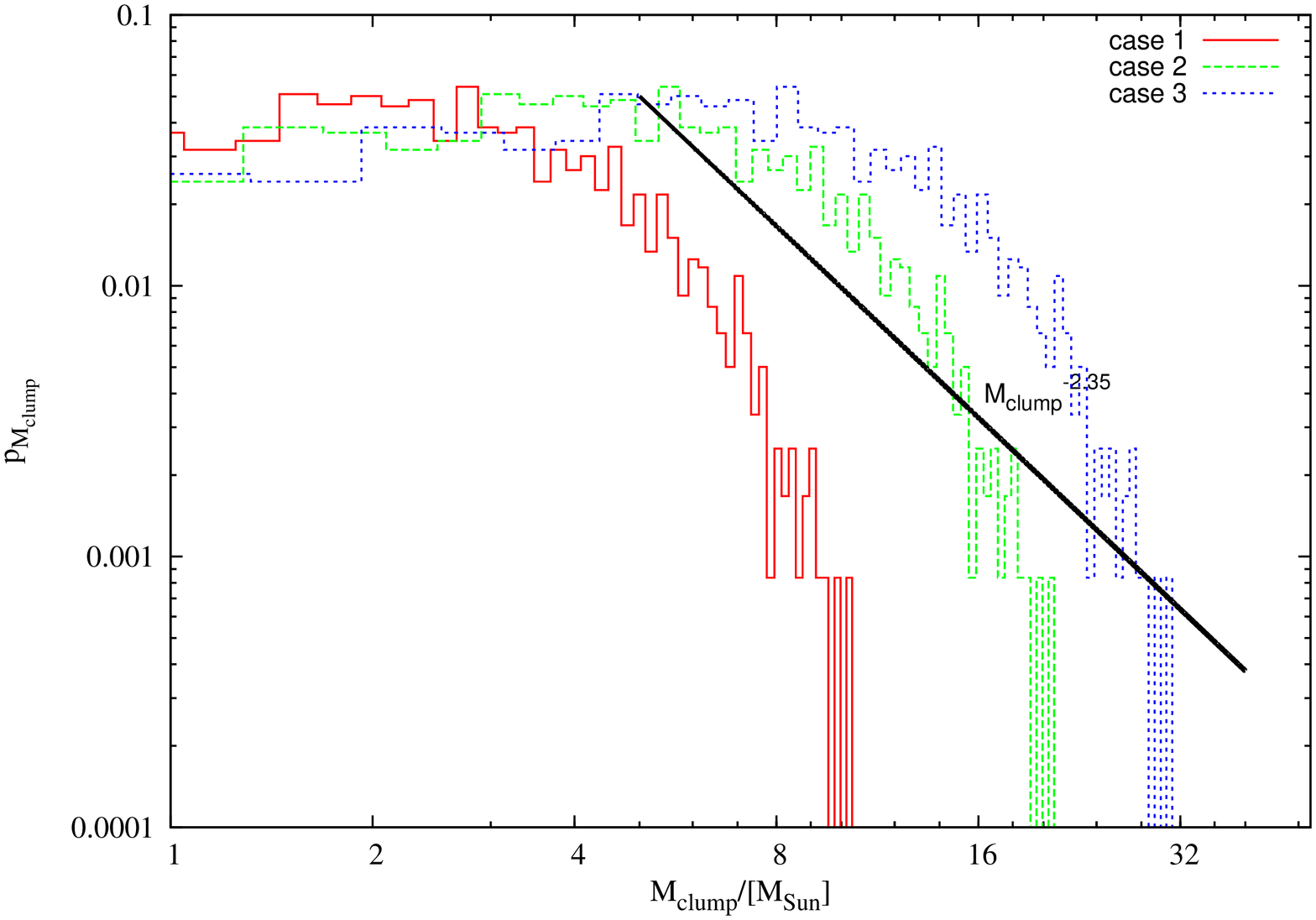}
  \caption{The initial mass-distribution for clumps detected in cases 1, 2 and 3 has been shown in these plots.}
\end{figure}

(d) \emph{Mass-spectrum} The distribution of masses for the cores detected in each realisation has been shown in Fig. 14. Across the three realisations, the  nature of the distribution of clump masses is largely similar, though the spectrum for cases 2 and 3 shifts right-ward relative to that for case 1. This is likely because, and as demonstrated in item (a) above, the clumps detected in these latter realisations are relatively warmer than those detected in case 1. It was also demonstrated in item (c) above that the fragmentation probably occurs on the scale of the local Jeans-length so that the turn-over of mass-spectra shown in Fig. 14 at the respective Jeans mass should be hardly surprising.  The shift in the turnover mass suggests that it is sensitive to the prevailing physical conditions. At this stage it is evident that sub-fragmentation of larger clumps into smaller cores is still likely underway as there is a significant population of massive fragments. However, even at this fairly early stage in their life-cycle, this initial mass-spectrum appears similar to that derived for cores detected in star-forming clouds (see e.g. Motte \emph{et al.} 1998; Nutter \& Ward-Thompson 2007; Enoch \emph{et al.} 2008), and is very well approximated by a power-law at the high-mass end. Using the peaks in respective histograms for $M_{clump}$ and $L_{clump}$, the condition for gravitational boundedness, $V^{2}\equiv\frac{2GM_{clump}}{L_{clump}}$, yields escape velocities on the order of $\sim 0.3$ km/s to $\sim 0.4$ km/s for the detected clumps in cases 1, 2 and 3, respectively. From the plots shown in Fig. 12 it is evident that the magnitude of gas-velocity dispersion for most, if not all the detected clumps is significantly larger than the escape velocity calculated above. The implication being, most clumps at this stage are likely unbound.

\section{Discussion }
Molecular clouds are replete with density structure that is often found to be filament-like with potential star-forming regions usually located at the junctions of these gas filaments (Schneider \emph{et al.} 2012). Filamentary clouds and potential star-forming clumps that form at the junctions of gas filaments  fragment further on the scale of local Jeans length to form prestellar cores. Interestingly though, cores, clumps and larger clouds all appear to follow empirical scaling relations first introduced by Larson (1981), which makes turbulence ubiquitously important. The fractal nature of molecular clouds suggests that clouds may not be collapsing universally instead, pockets of dense gas probably assembled via interactions between turbulent flows could collapse locally. Hydrodynamic instabilities are usually triggered due to shocks generated by strongly supersonic turbulent flows. One of the properties of these instabilities is their relatively short growth timescales that tends to fragment rapidly the shocked clouds while damping the gravitational instability (see eg. V{\' a}zquez-Semadeni \emph{et al.} 2007; Heitsch \emph{et al.} 2008 b; Anathpindika 2009). 

As demonstrated by the above authors, despite their rapid appearance  within a shocked cloud, fragments generated via growth of hydrodynamic instabilities are unlikely to be bound and  usually are transient by nature. Thus they are more likely to be either sheared apart by turbulent flows or perhaps merge with other neighbouring fragments.  On the other hand, confinement by relatively weaker shocks, or indeed, thermal pressure, yields radically different results as it is more likely to aid the growth of gravitational instability and produce more stable fragments that demonstrate a greater propensity towards a radial in-fall. However, irrespective of the type of confinement, gas-cooling appears to play a fundamental role in the process of fragmentation. A fragment (which could be a core), that cools efficiently becomes susceptible to gravitational collapse as it loses thermal support and as turbulence decays gradually. For the moment though we neglect the effects of magnetic field on dynamic stability. 

 Under appropriate physical conditions  the gravitational instability grows quickly and fragments the underlying gas cloud. The timescale of fragmentation, $\tau\equiv\frac{\lambda_{fast}}{a}$, where $a$ is the average sound-speed within the gas slab and $\lambda_{fast}$, length of the fastest growing unstable mode. If $E_{rad}$ is the excess energy radiated away by a pressure-confined cloud over its fragmentation timescale then, $E_{rad}\equiv\frac{p_{ext}}{n}$, and finally, the rate at which it loses energy, $\frac{dE_{rad}}{dt}\equiv n\Lambda - \Gamma$. Thus the timescale on which the gas cools, $\tau_{cool}\equiv\frac{E_{rad}}{dE_{rad}/dt}\equiv\frac{p_{ext}}{n^{2}\Lambda - n\Gamma}$. In a calculation such as this one, the cooling-timescale, of course, is determined by the functional dependence of the cooling function, $\Lambda$, on temperature. Eliminating time from the above expressions, we get an expression for the average sound-speed, $a$, within the fragmenting gas,
\begin{equation}
a\equiv\frac{n(n\Lambda - \Gamma)\lambda_{fast}}{p_{ext}}.
\end{equation}
 Equation (13) also shows that the average gas temperature attained by a gas-body, over a length scale, $\lambda_{fast}$, by radiating excess energy depends on the balance between the net heating and cooling. For the {\small GI} to grow gas within a cloud must cool to a sufficiently low temperature which makes cooling rate crucial for the fragmentation process.

 While the canonical Jeans analysis can account for the fragmentation process in the simplest case of an unconfined gas-body, pressure confinement adds a degree of complication. The gas-slab in this work evolved via a complex interplay of three instabilities, viz. the (i) thermal instability, (ii) shell-instability and (iii) gravitational instability. Fragmentation of the slab during its early stages is triggered by the confluence of the {\small TI} and the  shell-instability. Analytic arguments suggest, change in the confining pressure alone can possibly induce a change to the fate of a pressure-confined cloud. Equations (2) and (13) support this argument and  the scale of fragmentation is determined by the details of thermal physics. In the case of a cooling medium, the functional dependence of the cooling function on the gas-temperature must hold the key to determining the fate of the slab, or a cloud, in general.

\subsection{Thermal instability(TI) in Giant molecular clouds (GMCs)}
Depending on the magnitude of gravitational pressure in a {\small GMC}, it may be classified as pressure-confined or gravitationally bound. The magnitude of gravitational pressure in the former type of cloud is significantly smaller than its surface-pressure where-as in the latter, the gravitational pressure dominates the surface-pressure (e.g. Bertoldi \& Mc Kee 1992). It has been suggested that pressure-confined clouds may be considered part of the usual 3-phase model of the interstellar medium ({\small ISM}), while a more complex, 4-component model could be invoked for gravitationally bound {\small GMCs} (McKee 1995). The reason being, gravitational bound {\small GMCs} are not isobaric. However, it is now widely accepted that {\small GMCs}, irrespective of their gravitational state, are far from uniform and exhibit turbulent motions (e.g. Hughes \emph{et al.} 2013). In general, {\small GMCs} are therefore unlikely to be isobaric, though they are usually in approximate pressure equilibrium with the ambient {\small ISM}. So the suggested 4-component model may then be extended to {\small GMCs} without invoking the qualifying condition about gravitational boundedness.

The essence of this model is, {\small GMCs} are multi-phase media themselves and processes that regulate thermal-pressure in the diffuse {\small ISM} could very well regulate thermal-pressure in {\small GMCs} also. This model can then reconcile the fact that dense clumps in {\small GMCs} could possibly have thermal-pressure  comparable to, or even greater than that in typical HI clouds in the diffuse {\small ISM}. Now, the slab considered in numerical exercises discussed in this work is initially pressure-confined (see Table 1). It is rendered gravo-thermally unstable by continuous gas-cooling. Early growth of density perturbations in this slab disturb its initial isobaric nature. The rapid growth of these density perturbations cause redistribution of thermal pressure in the slab and the resulting fragmentation on the length-scale, $L_{fast}$, segregates the gas into the dense and a relatively tenuous phase, which we describe as the inter-clump medium. 

The resulting fragments continue to cool and during the process enhance their self-gravity. Though, a majority of the fragments in each realisation were still gravitationally unbound when calculations were terminated. Thermal pressure between the fragments and the inter-clump medium in which they float undergoes rapid redistribution during the course of evolution. Earlier, in \S 3.1 we demonstrated, fragments, soon after their formation, are at a relatively lower pressure relative to the inter-clump medium and some of them merge to form more contiguous objects. Eventually, the fragments acquire an approximate pressure-equilibrium with this inter-clump medium. This explains why the inter-clump medium is relatively warm compared to the dense-phase. Our conclusion is that a cooling molecular cloud with a non-uniform density distribution presents itself as a multi-phase medium and supports the hypothesis suggested by McKee (1995). It is also supported by observations that have demonstrated the tenuous nature of the inter-clump medium (e.g. Falgarone \& Puget 1988; Blitz 1991). More recently, Bailey \emph{et al.}(2014), have demonstrated that cores in molecular clouds are confined by a tenuous medium that is warm. Although we have succeeded in demonstrating the segregation of the density field, we could not follow the evolution of fragments to the stage when they would begin to collapse under self-gravity, but this is not the subject of this investigation.

\section{Conclusions}
 In our calculations we observed unstable modes first appearing on the slab-surface  via its interaction with the {\small ICM} confining it. These modes were associated with the simultaneous onset of the {\small TI} that induced fragmentation within the slab. This is true in all three realisations as is reflected by the higher growth-rate of the {\small TI} in comparison to that of the {\small GI}. We have demonstrated that molecular clouds are themselves 3-phase media and contain a dense phase, represented by clumps and cores, in approximate pressure equilibrium with the relatively tenuous inter-clump medium in which these clumps float and an intermediate unstable phase between these two phases.
Growth of these instabilities injected a velocity field within slab layers at large $k$ which then trickled down to small $k$. Gas in the dense phase continued to cool and fragmentation on the scale of the local Jeans length ensued. Interestingly, following a recent study of the dynamical properties of {\small MC}s across the M51, M33 and the {\small LMC} the authors suggested, pressure within a cloud and that confining it were correlated (Hughes \emph{et al.} 2013). This result is interesting since the induced velocity field, in a purely non-magnetic calculation, is a major contributor to the internal pressure and the magnitude of the externally confining pressure could affect the strength of this velocity field. 

The effects of gas-thermodynamics on physical properties of dense gas clumps is evident in this work. 

Clumps detected in the post-fragmentation slab are not only more massive, but predominantly unbound in the realisations where the temperature of the initial volume of gas deviated strongly from its equilibrium value. These clumps formed on a longer time-scale and also had a relatively large velocity dispersion. Sub-fragmentation 
of these clumps to form smaller cores is possible once they acquire sufficient mass, or conversely, have cooled sufficiently. Irrespective of this difference, the mass-spectrum qualitatively resembles the mass-spectrum for clumps and/or cores found in typical star-forming clouds. This re-emphasises the suggestion that the core mass spectrum is probably seeded by the core-formation process itself (e.g. Heitsch et al. 2008 a).  

A comparison of the results obtained from this work with those in the literature for an isothermal slab demonstrate the effects of gas thermodynamics on the evolution of the slab. Although the outcome is broadly consistent, i.e., the slab fragments irrespective of its thermodynamic state, the timescale and the length-scale on which this happens is significantly different in either type. Unlike the isothermal slab,  the timescale of fragmentation for a dynamically cooling slab is determined by the cooling function as demonstrated the expression for the cooling timescale, $\tau_{cool}$, defined by Eqn. (13). Although sufficient to demonstrate the ongoing dynamic processes, the assumption of an isothermal gas will most probably bias results in a direction dictated by the initial choice of temperature; see for instance Bonnell \emph{et al.}(2006). A self-consistent calculation of gas temperature with a cooling function permits us to demonstrate the effect of prevailing physical conditions on various properties of clumps as was demonstrated in the previous section.

Prestellar cores are found at sub-parsec scale and more recent observations of cores have shown that gas within potential star-forming cores is cold at only a few Kelvin and is pervaded by a weak velocity field, often referred to as micro-turbulence. Low-magnitude velocity fields have been observed in both, bound prestellar cores as well as unbound cores (see e.g. Motte \emph{et al.} 1998; Jijina \emph{et al.} 1999; Hatchell \emph{et al.} 2005; Nutter \& Ward-Thompson 2007; Enoch \emph{et al.} 2008; among a number of other authors working on the subject). It is commonly believed, this velocity field within cores is introduced during their natal phase itself while they are yet accreting gas from their parent clouds. Although this is quite likely, results from this work show that a turbulent velocity field is probably first injected in virgin clouds via interactions with the confining {\small ICM}, followed by growth of other dynamical instabilities such as the {\small TI}. 

From our test simulations it appears that cores and their parent clumps form on different timescales which makes the cooling timescale, $\tau_{cool}$, of molecular gas within clumps, crucial. We have seen that interaction with a warmer ICM in the case of a strongly confined slab introduces a relatively  strong velocity field. So although the initial fragmentation in these realisations happens on a relatively short timescale, fragments are unbound and some of them do merge to form larger objects. Thus, while a turbulent velocity field need not necessarily delay fragmentation, the formation of potential star-forming clumps is likely to take longer.

Using a relatively simple prescription of the cooling function we have demonstrated the possible variation in properties of cores in response to changes in the externally confining pressure. We have also shown that a confluence of various dynamic instabilities could possibly seed a weak velocity field in dense clumps of gas. Thus it is clear that dynamic processes preceding the prestellar cores are responsible for their observed physical properties. The apparent importance of the thermodynamic properties of gas in the process of fragmentation is consonant with the canonical Jeans analysis. Although in the present work we demonstrated this using a relatively simple cooling function, the real gas dynamics is  much more complex and must be accounted for.

 Pure hydrodynamic models such as those discussed here must be coupled with the appropriate molecular chemistry so that evolution of molecular gas can be understood better. Such integrated models can then be used to gauge the effects of cooling of molecular gas on its dynamical stability and study the onset of fragmentation, in other words, formation of dense clumps and cores.  While reproducing the typical nature of the distribution of clump masses, apart from other important physical properties, we have here demonstrated that the seeds of core-formation are introduced during the process of dynamical fragmentation itself and of course, the rate with which molecular gas cools likely determines the efficiency of core-formation.  We have also demonstrated that the weak velocity field often found within cores is probably injected during the fragmentation phase itself. It is also apparent that turbulence in virgin clouds can appear without any artificial injection and is probably seeded by the growth of various dynamical instabilities. For a further understanding of the process of core-formation and the evolution of cores however, we  must supplement  purely hydrodynamic models of {\small MC}s with the appropriate molecular chemistry.

\begin{acknowledgements}
The author gratefully acknowledges an anonymous referee for a helpful report that helped improve the original manuscript. 
\end{acknowledgements}


\begin{thebibliography}{}
\bibitem[\protect\citeauthoryear{Alves}{2007}]{b130}Alves, J., Lombardi, M \& Lada, C. J., 2007, A\&A, 462, L17
\bibitem[\protect\citeauthoryear{Anathpindika1}{2009}]{b1} Anathpindika, S., 2009, A\&A, 504, 437
\bibitem[\protect\citeauthoryear{Anathpindika2}{2011}]{b2} Anathpindika, S., 2011, New Astronomy, 16, 477
\bibitem[\protect\citeauthoryear{Anathpindika3}{2011}]{b3} Anathpindika, S., 2013, New Astronomy, 18, 6
\bibitem[\protect\citeauthoryear{Andre}{2007}]{b135}Andr{\' e}, Ph.,
Belloche, A., Motte, F \& Peretto, N., 2007, A\&A, 472, 519
\bibitem[\protect\citeauthoryear{Andre}{2010}]{b5}Andr{\' e}, Ph., Men'shchikov, A., Bontemps, S \emph{et al.}, 2010, A\&A, 518, L102
\bibitem[\protect\citeauthoryear{Audit}{2010}]{b7} Audit, E \& Hennebelle, P., 2010, A\&A, 511, 76
\bibitem[\protect\citeauthoryear{BaLLEST}{2011}]{b105} Ballesteros-Paredes, J.,V{\' a}zquez-Semadeni, E., Gazol, Adriana, Hartmann, L., Heitsch, F \& Colin, P., 2011, MNRAS, 416, 1436
\bibitem[\protect\citeauthoryear{Banerjee}{2009}]{b107} Banerjee, R \emph{et al.}, 2009, MNRAS, 398, 1082
\bibitem[\protect\citeauthoryear{Basu}{2014}]{b128} Bailey, N., Basu, S \& Caselli, P., 2014, \emph{to appear in ApJ}, astroph. 1410.4425
\bibitem[\protect\citeauthoryear{Bate}{2009}]{b8} Bate, M., 2009, MNRAS, 392, 590
\bibitem[\protect\citeauthoryear{Bertoldi}{2009}]{b8} Bertoldi, F \& McKee, C. F., 1992, ApJ, 395, 140
\bibitem[\protect\citeauthoryear{Beuther}{2013}]{b54}Beuther, H., Linz, H., Tackenberg, J., Henning, Th \emph{et al.}, 2013, A\&A, 553, A115
\bibitem[\protect\citeauthoryear{Blitz}{1991}]{b109} Blitz, L.,1991, \emph{The physics of star-formation and early stellar-evolution}, Eds. Lada, C. \& Kylafis, N.; Dordrecht, Kluwer, 3
\bibitem[\protect\citeauthoryear{Bonnell}{2006}]{b9} Bonnell, I., Clarke, C \& Bate, M., 2006, MNRAS, 368, 1296
\bibitem[\protect\citeauthoryear{Bonnell}{2013}]{b10} Bonnell, I., Dobbs, C \& Smith, R. J., 2013, MNRAS.tmp.782B
\bibitem[\protect\citeauthoryear{Bonnor}{1956}]{b58} Bonnor, W., 1956, MNRAS, 116, 351
\bibitem[\protect\citeauthoryear{cLARKE}{1999}]{b12} Clarke, C. J., 1999, MNRAS, 307, 328
\bibitem[\protect\citeauthoryear{cLARKE}{2007}]{b120} Clarke, C. J. \& Carswell, R. F., 2007, \emph{Principles of Astrophysical Fluid Dynamics, pgs. 146-47}, Cambridge University Press, UK.
\bibitem[\protect\citeauthoryear{cLARKE}{1999}]{b106} Chabrier, G., 2003, ApJL, 586, L133
\bibitem[\protect\citeauthoryear{cURTIS}{2010}]{b13} Curtis, E., Richer, J \& Buckle, J., 2010, MNRAS, 401, 455
\bibitem[\protect\citeauthoryear{Dib}{2005}]{b103} Dib, S \& Burkert, A., 2005, ApJ, 630, 238
\bibitem[\protect\citeauthoryear{Dobbs}{2012}]{b113} Dobbs, C., Pringle, J. E. \& Burkert, A., 2012, MNRAS, 425, 2157
\bibitem[\protect\citeauthoryear{Donovan}{2013}]{b144}Donovan Meyer, J., Koda, J., Momose, R., Mooney, T \emph{et al.}, 2013, ApJ, 772, 107
\bibitem[\protect\citeauthoryear{Duarte}{2010}]{b116} Duarte-Cabral \emph{et al.}, 2010, A\&A, 519, 27
\bibitem[\protect\citeauthoryear{Eisenstein}{1998}]{b63}Eisenstein, \& Hut, B., 1998, ApJ, 498, 137
\bibitem[\protect\citeauthoryear{Elmegreen}{1978}]{b58} Elmegreen, B \& Elmegreen, D., 1978, ApJ, 220, 1051
\bibitem[\protect\citeauthoryear{Elmegreen}{1996}]{b142} Elmegreen, B. G. \& Falgarone, E., 1996, ApJ, 471, 816
\bibitem[\protect\citeauthoryear{Enoch}{2008}]{b14} Enoch, M., Evans, N. J. II, Sargent, A. I., Glenn, J., Rosolowsky, E \& Myers, P. C., 2008, ApJ, 684, 1240
\bibitem[\protect\citeauthoryear{Falgarone}{1988}]{b110} Falgarone, E. \& Puget, J., P., 1988, \emph{Galactic and Extragalactic star-formation}; Eds. Pudritz, R \& Fich, M.; Dordrecht, Kluwer, 195
\bibitem[\protect\citeauthoryear{Falgarone}{2004}]{b56}Falgarone, E., Hily-Blant, P \& Levrier, F., 2004, Ap\&SS, 292, 285
\bibitem[\protect\citeauthoryear{fALGARONE}{2008}]{b15} Falgarone, E., Pety, J \& Hily-Blant, P., 2009, A\&A, 507, 355
\bibitem[\protect\citeauthoryear{Federrath}{2008}]{b16} Federrath, F., Klessen, R \& Schmidt, W., 2008, ApJ, 688, L79
\bibitem[\protect\citeauthoryear{Federrath}{2010}]{b134}Federrath, C., Roman-Duval, J., Klessen, R., Schmidt, W \& Mac Low, M. M., 2010, A\&A, 512, A81 
\bibitem[\protect\citeauthoryear{Federrath}{2011}]{b154}Federrath, C., Sur, S., Schleicher, D., Banerjee, R \& Klessen, R., 2011, ApJ, 731, 62
\bibitem[\protect\citeauthoryear{Federrath}{2012}]{b131}Federrath, C \& Klessen, R., 2012, ApJ, 761, 156
\bibitem[\protect\citeauthoryear{Federrath}{2013}]{b155}Federrath, C \& Klessen, R., 2012, ApJ, 763, 51
\bibitem[\protect\citeauthoryear{Federrath}{2012}]{b133}Federrath, C., 2013, MNRAS, 436, 1245 
\bibitem[\protect\citeauthoryear{Federrath}{2012}]{b153} Federrath, C., Schr$\ddot{\mathrm{o}}$n, M., Banerjee, R \& Klessen, R., 2014, ApJ, 790, 128
\bibitem[\protect\citeauthoryear{Field}{1965}]{b101} Field, G. B., ApJ, 1965, 142, 531
\bibitem[\protect\citeauthoryear{Gazol}{2013}]{b138} Gazol, A\& Kim, J., 2013, ApJ, 765, 49
\bibitem[\protect\citeauthoryear{Glover}{2010}]{b139} Glover, S.C. O., Federrath, C., Mac Low, M. M. \& Klessen, R., 2010, MNRAS, 404, 2
\bibitem[\protect\citeauthoryear{Goodman}{1998}]{b151}Goodman, A., Barranco, J., A., Wilner, D. \& Heyer, M., 1998, ApJ, 504, 223
\bibitem[\protect\citeauthoryear{Graves}{2010}]{b118} Graves \emph{et al.}, 2010, MNRAS, 409, 1412
\bibitem[\protect\citeauthoryear{Hartmann}{2002}]{b17} Hartmann, L., 2002, ApJ, 578, 914
\bibitem[\protect\citeauthoryear{hATCHELL}{2005}]{b18} Hatchell, J., Richer, J., Fuller, G., Qualtrough, C., Ladd, C \& Chandler, C., 2005, A\&A, 440, 151
\bibitem[\protect\citeauthoryear{Hily}{2009}]{b150}Hily-Blant, P \& Falgarone, E., 2009, A\&A, 500, L29
\bibitem[\protect\citeauthoryear{hENNEBELLE}{2008}]{b19} Hennebelle, P \& Chabrier, G., 2008, ApJ, 684, 395
\bibitem[\protect\citeauthoryear{Hennebelle}{Hennebelle}{2012}]{b20} Hennebelle, P \& Falgarone, E., 2012, A\&ARv, 20, art. id. 55
\bibitem[\protect\citeauthoryear{Heithausen}{1998}]{b54} Heithausen, A., Bensch, F., Stutzki, J., Falgarone, E \& Parnis, J., 1998, A\&A, 331, L65
\bibitem[\protect\citeauthoryear{Heitsch}{2008}]{b107}Heitsch, F., Hartmann, L., Slyz, A., Devreindt, J. \& Burkert, A., 2008, ApJ, 674, 316  (Heitsch et al. a)
\bibitem[\protect\citeauthoryear{Heitsch}{2008}]{b21} Heitsch, F., Hartmann, L \& Burkert, A., 2008, ApJ, 683, 786 (Heitsch et al. b)
\bibitem[\protect\citeauthoryear{Heyer}{2009}]{b57}Heyer, M., Krawczyk, C., Duval, J \& Jackson, J., 2009, ApJ, 699, 1092
\bibitem[\protect\citeauthoryear{Hobbs}{2013}]{b22} Hobbs, A., Read, J., Power, C \& Cole, D., MNRAS, 2013, 434, 1849
\bibitem[\protect\citeauthoryear{Hubber}{2006}]{b115}Hubber, D., Goodwin, S \& Whitworth, A. P., 2006, A\&A, astroph0512247
\bibitem[\protect\citeauthoryear{Hubber}{2011}]{b23}Hubber, D., Batty, C., McLeod, A \& Whitworth, A., 2011, A\&A, 529, 28
\bibitem[\protect\citeauthoryear{Hughes}{2013}]{b65}Hughes, A., Meidt, S., Colombo, D., Schinnerer, E., Pety, J., Leroy, A., Dobbs, C., Garcia-Burillo, S. \emph{et al.}, 2013, to appear in ApJ, astroph 1309.3453

\bibitem[\protect\citeauthoryear{Hunter1}{1979}]{b24}Hunter, J.(Jr), 1979, ApJ, 233, 946
\bibitem[\protect\citeauthoryear{Hunter2}{1982}]{b25}Hunter, J.(Jr) \& Fleck, R, 1982, ApJ, 256, 505

\bibitem[\protect\citeauthoryear{Inutsuka}{1997}]{b53}Inutsuka, S \& Miyama, S., 1997, ApJ, 480, 681
\bibitem[\protect\citeauthoryear{Jeans}{1928}]{b100} Jeans, J. H., \emph{Astronomy and Cosmogony}. Cambridge : Cambridge University Press; reprinted by Dover
\bibitem[\protect\citeauthoryear{Jijina}{1999}]{b11} Jijina, J., Myers, P \& Adams, F., 1999, ApJSS, 125, 161
\bibitem[\protect\citeauthoryear{Kainulainen}{2009}]{b26}Kainulainen, J., Beuther, H., Henning, T., Henning, T \& Plume, R., 2009, ApJ, 508, L35
\bibitem[\protect\citeauthoryear{Kainulainen}{2014}]{b140}Kainulainen, J., Federrath, C \& Henning, Th., 2014, Science, 344, 183
\bibitem[\protect\citeauthoryear{Kim}{2005}]{b60} Kim, J \& Ryu, D., 2005, ApJ, 630, L45
\bibitem[\protect\citeauthoryear{Kirk}{2007}]{b27} Kirk, J. M., Ward-Thompson, D \& Andr{\' e}, Ph., 2007, MNRAS, 375, 843
\bibitem[\protect\citeauthor{Kitsionas}{2009}]{145}Kitsionas, S., Federrath, C., Klessen, R., S., Schmidt, W.,Price, D., J., Dursi, L., J \emph{et al.}, 2009, A\&A, 508, 541 
\bibitem[\protect\citeauthoryear{Kolmogorov}{1941}]{b60} Kolmogorov, A., 1941, Proc. of R. Soc. London, Series A 434, Reprint 1991
\bibitem[\protect\citeauthoryear{Koyoma}{2002}]{b28}Koyoma, H \& Inutsuka, S., 2002, ApJ, 564, L97
\bibitem[\protect\citeauthoryear{Koyoma}{2004}]{b158}Koyoma, H \& Inutsuka, S., 2004, ApJ, 602, L25
\bibitem[\protect\citeauthoryear{Klessen}{2000}]{b29} Klessen, R. \& Burkert, A., 2000, ApJ, 128, 287
\bibitem[\protect\citeauthoryear{Klessen}{2005}]{b30} Klessen, R., Ballesteros-Paredes, J., V{\' a}zquez-Semadeni, E \& Roman-Duval, C., 2005, ApJ, 620, 786
\bibitem[\protect\citeauthoryear{Kramer}{1998}]{b55} Kramer, C., Stutzki, J., Rohrig, R \& Corneliussen, U., 1998, A\&A, 329, 249
\bibitem[\protect\citeauthoryear{Kritsuk}{2011}]{b156}Kritsuk, A., Norman, M \& Wagner, R., ApJ, 727, L20
\bibitem[\protect\citeauthoryear{Larson}{1981}]{b28}Larson, R., 1981, MNRAS, 194, 809
\bibitem[\protect\citeauthoryear{Ledoux}{1951}]{b58} Ledoux, P., 1951, Ann. Astrophysics, 14, 438
\bibitem[\protect\citeauthoryear{Lubow}{1993}]{b57} Lubow, S \& Pringle, S., 1993, MNRAS, 263, 701
\bibitem[\protect\citeauthoryear{Mac Low}{2004}]{b33} Mac Low, M. M. \& Klessen, R., 2004, Rev. of Mod. Phys., 76, 125
\bibitem[\protect\citeauthoryear{McKee}{1995}]{b157} McKee, C. F., 1995, \emph{The Physics of ISM \& IGM, ASPC}, 80, 292
\bibitem[\protect\citeauthoryear{Miesch}{1995}]{b53}Miesch, M \& Scalo, J., 1995, ApJ, L27
\bibitem[\protect\citeauthoryear{Monaghan}{1997}]{b125}Monaghan, J., JCoPh., 136, 298
\bibitem[\protect\citeauthoryear{Motte}{1998}]{b34}Motte, F., Andr{\' e}, Ph \& Neri, R., 1998, A\&A, 336, 150
\bibitem[\protect\citeauthoryear{nUTTER}{2007}]{b38} Nutter, D \& Ward-Thompson, D., 2007, MNRAS, 374, 1413
\bibitem[\protect\citeauthoryear{Padoan}{1997}]{b59} Padoan, P., Jones, B \& Nordlund, $\mathring{A}$, 1997, ApJ, 474, 730
\bibitem[\protect\citeauthoryear{Padoan}{1999}]{b35} Padoan, P., Nordlund, $\mathring{A}$ \& Jones, B., 1999, MNRAS, 288, 145
\bibitem[\protect\citeauthoryear{Price}{2007}]{b36} Price, D., 2007, PASA, 24(30), 159
\bibitem[\protect\citeauthoryear{Price}{2008}]{b36} Price, D., 2008, JCoPhys, 227, 10040
\bibitem[\protect\citeauthoryear{Price}{2008}]{b136} Price, D. \& Bate, M., 2008, MNRAS, 386, 1820
\bibitem[\protect\citeauthoryear{Price}{2009}]{b137} Price, D. \& Bate, M., 2009, MNRAS, 398, 33
\bibitem[\protect\citeauthoryear{Price}{2010}]{b37} Price, D \& Federrath, C., 2010, MNRAS, 406, 1659
\bibitem[\protect\citeauthoryear{Price}{2011}]{b132} Price, D., Federrath, C. \& Brunt, C. M., 2011, ApJ, 727, L21
\bibitem[\protect\citeauthoryear{Ragan}{2012}]{b39} Ragan, S., Heitsch, F., Bergin, E \& Wilner, D., 2012, ApJ, 746, 174
\bibitem[\protect\citeauthoryear{Ripple}{2013}]{b40} Ripple, F., Heyer, M., Gutermuth, R., Snell, R \& Brunt, C., 2013, MNRAS, 431, 1296
\bibitem[\protect\citeauthoryear{Roman-Duval}{2010}]{b41} Roman-Duval, J., Jackson, J., Heyer, M., Rathborne, J., Simon, R., 2010, ApJ, 723, 492
\bibitem[\protect\citeauthoryear{Schmidt}{2010}]{b42} Schmidt, W., Kern, S.A.W., Federrath, C \& Klessen, R., 2010, A\&A, 516, 25
\bibitem[\protect\citeauthoryear{Schneider}{2011}]{b43}Schneider, N., Bontemps, S., Simon, R., Ossenkopf, V., Federrath, C., Klessen, R. S., Motte, F., Andr{\' e}, Ph., Stutjki, J \& Brunt, C., 2011, A\&A, A1, 529
\bibitem[\protect\citeauthoryear{Schneider}{2012}]{b141}Schneider, N., Csengeri, T., Hennemann, M., Motte, F., Didelon, P., Federrath, C. \emph{et al.}, 2012, A\&A, 540, L11
\bibitem[\protect\citeauthoryear{Schneider}{2013}]{b142}Schneider, N., Andr{\' e}, Ph., K$\ddot{\mathrm{o}}$nyves, V., Bontemps, S., Motte, F., Federrath, C., Ward-Thompson, D \emph{et al.}, 2013, ApJ, 766, L17
\bibitem[\protect\citeauthoryear{Tohline}{1987}]{b44} Tohline, J., Bodenheimer, P \& Christodoulou, D., 1987, ApJ, 322, 787
\bibitem[\protect\citeauthoryear{Truelove}{1997}]{b} Truelove, J. K., Klein, R. I., McKee, C. F., Holliman, J., H.II, Howell, L. K \& Greenhough, J. A., 1997, ApJ, 489,L179
\bibitem[\protect\citeauthoryear{Tsitali}{2014}]{b63}Tsitali, A. E., Belloche, A., Garrod, R. T., Parise, B \& Menten, K. M., 2014, \emph{to appear in A\& A}, astroph 1411.3741
\bibitem[\protect\citeauthoryear{Semadeni}{1994}]{b45}V{\' a}zquez-Semadeni, E., 1994, ApJ, 423, 681
\bibitem[\protect\citeauthoryear{Semadeni}{2003}]{b146}V{\' a}zquez-Semadeni, E., Passot, T \& Pouquet, A., 1996, ApJ, 473, 881
\bibitem[\protect\citeauthoryear{Semadeni}{2003}]{b135}V{\' a}zquez-Semadeni, E., Ballesteros-Paredes, J \& Klessen, R., 2003, ApJ, 585, L131 
\bibitem[\protect\citeauthoryear{Semadeni}{2007}]{b46}V{\' a}zquez-Semadeni, E., G{\' o}mez, G., Jappsen, A. K., Ballesteros-Paredes, J, Gonz{\' a}lez, R. \& Klessen, R., 2007, ApJ, 657, 87
\bibitem[\protect\citeauthoryear{Semadeni}{2008}]{b104}V{\' a}zquez-Semadeni, E., Gonz{\' a}lez, Ricardo, F., Ballesteros-Paredes, J., Gazol, A \& Kim, J., 2008, MNRAS, 390, 769
\bibitem[\protect\citeauthoryear{Vishniac}{1983}]{b47}Vishniac, E., 1983, ApJ, 274, 152
\bibitem[\protect\citeauthoryear{Ward-Thompson}{2007}]{b48} Ward-Thompson, D., Di FRancesco, J., Hatchell, J \emph{et. al.}, 2007, PASP, 119, 855 
\bibitem[\protect\citeauthoryear{wHITWORTH1}{1981}]{b49} Whitworth, A.,1981, MNRAS, 195, 967
\bibitem[\protect\citeauthoryear{wHITWORTH2}{1994}]{b50} Whitworth, A., Bhattal, A., Chapman, S., Disney, M \& Turner, J., 1994, A\&A, 290, 421
\bibitem[\protect\citeauthoryear{wOLFIRE}{1995}]{b51} Wolfire, M., McKee, C., Hollenbach, D \& Tielens, A.G.G.M., 1995, ApJ, 453, 673
\bibitem[\protect\citeauthoryear{Yamada\&Nishi}{1998}]{b52} Yamada, M \& Nishi, R., 1998, ApJ, 505, 148 


\end{thebibliography}
\end{document}